\documentclass[journal]{IEEEtran}
\usepackage{amsmath}
\usepackage{graphicx}
\usepackage[hidelinks]{hyperref}
\usepackage{amsthm}
\usepackage{multirow}
\usepackage{booktabs}
\usepackage{algorithm}
\usepackage{algorithmic}
\usepackage{color}
\usepackage{amsfonts}
\usepackage{ulem}
\usepackage{float}
\usepackage{bm}
\usepackage{caption}

\ifCLASSOPTIONcompsoc
  \usepackage[nocompress]{cite}
\else
  \usepackage{cite}
\fi
\ifCLASSINFOpdf
\else
\fi
\begin{document}
\setlength{\textfloatsep}{-1pt}
% paper title
% Titles are generally capitalized except for words such as a, an, and, as,
% at, but, by, for, in, nor, of, on, or, the, to and up, which are usually
% not capitalized unless they are the first or last word of the title.
% Linebreaks \\ can be used within to get better formatting as desired.
% Do not put math or special symbols in the title.
\title{TEA: A Sequential Recommendation Framework via Temporally Evolving Aggregations}
%
%
% author names and IEEE memberships
% note positions of commas and nonbreaking spaces ( ~ ) LaTeX will not break
% a structure at a ~ so this keeps an author's name from being broken across
% two lines.
% use \thanks{} to gain access to the first footnote area
% a separate \thanks must be used for each paragraph as LaTeX2e's \thanks
% was not built to handle multiple paragraphs
%
%
%\IEEEcompsocitemizethanks is a special \thanks that produces the bulleted
% lists the Computer Society journals use for "first footnote" author
% affiliations. Use \IEEEcompsocthanksitem which works much like \item
% for each affiliation group. When not in compsoc mode,
% \IEEEcompsocitemizethanks becomes like \thanks and
% \IEEEcompsocthanksitem becomes a line break with idention. This
% facilitates dual compilation, although admittedly the differences in the
% desired content of \author between the different types of papers makes a
% one-size-fits-all approach a daunting prospect. For instance, compsoc 
% journal papers have the author affiliations above the "Manuscript
% received ..."  text while in non-compsoc journals this is reversed. Sigh.

\author{Zijian Li, Ruichu Cai$^\star$,~\IEEEmembership{Senior Member,~IEEE,}
	Fengzhu Wu, Sili Zhang, Hao Gu, Yuexing Hao, Yuguang Yan$^\star$% <-this % stops a space
	\IEEEcompsocitemizethanks{
		\IEEEcompsocthanksitem Zijian Li is with the School of Computer, Guangdong University of Technology, Guangzhou China, 510006.
		E-mail: leizigin@gmail.com
		\IEEEcompsocthanksitem Ruichu Cai is with the School of Computer, Guangdong University of Technology and Guangdong Provincial Key Laboratory of Public Finance and Taxation with Big Data Application, Guangzhou China, 510006.
		E-mail: cairuichu@gmail.com
		\IEEEcompsocthanksitem Fengzhu Wu is with the School of Computer, Guangdong University of Technology, Guangzhou China, 510006.
		E-mail: fzwu97@gmail.com
		\IEEEcompsocthanksitem Sili Zhang is with the School of Computer, Guangdong University of Technology, Guangzhou China, 510006. E-mail: zhangsili1260@gmail.com
		\IEEEcompsocthanksitem Hao Gu is with Tencent Technology (SZ) Co., Ltd. E-mail: nickgu@tencent.com
		\IEEEcompsocthanksitem Yuexing Hao is with Cornell University. E-mail:yh727@cornell.edu
		\IEEEcompsocthanksitem Yuguang Yan is with the School of Computer, Guangdong University of Technology, Guangzhou China, 510006. E-mail: ygyan@gdut.edu.cn
		}% <-this % stops an unwanted space
	\thanks{Manuscript received XX; revised XX; accepted XX. Date of publication XX XX, 2019; date of current version XX XX, 2019. This research was supported in part by National Key R\&D Program of China (2021ZD0111501),  National Science Fund for Excellent Young Scholars (62122022), Natural Science Foundation of China (61876043, 61976052) and Guangdong Provincial Science and Technology Innovation Strategy Fund (2019B121203012).   (\emph{Corresponding author: Ruichu Cai and Yuguang Yan.})
	}
}

% note the % following the last \IEEEmembership and also \thanks - 
% these prevent an unwanted space from occurring between the last author name
% and the end of the author line. i.e., if you had this:
% 
% \author{....lastname \thanks{...} \thanks{...} }
%                     ^------------^------------^----Do not want these spaces!
%
% a space would be appended to the last name and could cause every name on that
% line to be shifted left slightly. This is one of those "LaTeX things". For
% instance, "\textbf{A} \textbf{B}" will typeset as "A B" not "AB". To get
% "AB" then you have to do: "\textbf{A}\textbf{B}"
% \thanks is no different in this regard, so shield the last } of each \thanks
% that ends a line with a % and do not let a space in before the next \thanks.
% Spaces after \IEEEmembership other than the last one are OK (and needed) as
% you are supposed to have spaces between the names. For what it is worth,
% this is a minor point as most people would not even notice if the said evil
% space somehow managed to creep in.

% The paper headers
\markboth{IEEE Transactions on Neural Networks and Learning Systems,~submitted}%
% \markboth{}%
{Ruichu Cai \MakeLowercase{\textit{et al.}}: TEA: A Sequential Recommendation Framework via Temporally Evolving Aggregation}
% The only time the second header will appear is for the odd numbered pages
% after the title page when using the twoside option.
% 
% *** Note that you probably will NOT want to include the author's ***
% *** name in the headers of peer review papers.                   ***
% You can use \ifCLASSOPTIONpeerreview for conditional compilation here if
% you desire.

% The publisher's ID mark at the bottom of the page is less important with
% Computer Society journal papers as those publications place the marks
% outside of the main text columns and, therefore, unlike regular IEEE
% journals, the available text space is not reduced by their presence.
% If you want to put a publisher's ID mark on the page you can do it like
% this:
%\IEEEpubid{0000--0000/00\$00.00~\copyright~2015 IEEE}
% or like this to get the Computer Society new two part style.
%\IEEEpubid{\makebox[\columnwidth]{\hfill 0000--0000/00/\$00.00~\copyright~2015 IEEE}%
%\hspace{\columnsep}\makebox[\columnwidth]{Published by the IEEE Computer Society\hfill}}
% Remember, if you use this you must call \IEEEpubidadjcol in the second
% column for its text to clear the IEEEpubid mark (Computer Society jorunal
% papers don't need this extra clearance.)

% use for special paper notices
%\IEEEspecialpapernotice{(Invited Paper)}

% for Computer Society papers, we must declare the abstract and index terms
% PRIOR to the title within the \IEEEtitleabstractindextext IEEEtran
% command as these need to go into the title area created by \maketitle.
% As a general rule, do not put math, special symbols or citations
% in the abstract or keywords.
\IEEEtitleabstractindextext{%
\begin{abstract}
Sequential recommendation aims to \textcolor{black}{choose the most suitable items} for a user at a specific timestamp given historical behaviors.
\textcolor{black}{Existing methods usually model the user behavior sequence based on transition-based methods like Markov Chain.
% assume that the user behavior sequence is evolving periodically and model it with the transition-based methods like Markov Chain. 
However, these methods also implicitly assume that the users are independent of each other without considering the influence between users. In fact, this influence plays an important role in sequence recommendation since the behavior of a user is easily affected by others. Therefore, it is desirable to aggregate both user behaviors and the influence between users, which are evolved temporally and involved in the heterogeneous graph of users and items. }
% However, these methods assume that the relationship of users and items is fixed
% and the heterogeneous graph is static,
% which cannot reflect the dynamic influence of temporal interactions between users and items in a sequential recommendation system.
In this paper, we incorporate dynamic user-item heterogeneous graphs to propose a novel sequential recommendation framework. As a result, the historical behaviors as well as the influence between users can be taken into consideration. To achieve this, we firstly formalize sequential recommendation as a problem to estimate conditional probability given temporal dynamic heterogeneous graphs and user behavior sequences.
After that, we exploit the conditional random field to aggregate the heterogeneous graphs and user behaviors for probability estimation,
and employ the pseudo-likelihood approach to derive a tractable objective function.
% By doing this,
% both the temporal dynamic user-item interactions and the historical behaviors are leveraged in the energy functions in our model.
Finally, we provide scalable and flexible implementations of the proposed framework. Experimental results on three real-world datasets not only demonstrate the effectiveness of our proposed method but also provide some insightful discoveries on the sequential recommendation.
\end{abstract}
% Recent years have witnessed tremendous interest in recommendation combined with the social networks. Two types of information should be taken into account: the historical behaviors sequence of users and the dynamic heterogeneous graph between the users and the items. However, existing methods rarely leverage the aforementioned information simultaneously. Hence, it is desired to devise a unified principle to guide how to aggregate these two kinds of information. In this paper, we propose a novel social recommendation framework via aggregating the historical user behaviors and the dynamic heterogeneous propagation. First, we formalize the goal of social recommendation as a conditional probability estimation problem given the user behavior sequence and the temporally evolving heterogeneous graphs sequence. Second, we derive the objective function with the help of conditional random field. We find that the objective function is composed of the unary energy function and the pairwise energy function, which correspond to the historical behaviors aggregation and the temporally evolving user aggregation and item aggregation. Furthermore, we employ the pseudo-likelihood method to derive a tractable objective function. Finally, we provide a scalable, flexible and practical implementation of these three components. Experimental results on three real-world data sets not only demonstrate the effectiveness of our proposed method but also provide some insightful discoveries on social recommendation.
% Note that keywords are not normally used for peerreview papers.
}

% make the title area
\maketitle
% To allow for easy dual compilation without having to reenter the
% abstract/keywords data, the \IEEEtitleabstractindextext text will
% not be used in maketitle, but will appear (i.e., to be "transported")
% here as \IEEEdisplaynontitleabstractindextext when the compsoc 
% or transmag modes are not selected <OR> if conference mode is selected 
% - because all conference papers position the abstract like regular
% papers do.
\IEEEdisplaynontitleabstractindextext
\IEEEpeerreviewmaketitle

% \IEEEraisesectionheading{}
\section{Introduction}

\begin{figure*}[htbp]
	
	\includegraphics[width=\textwidth]{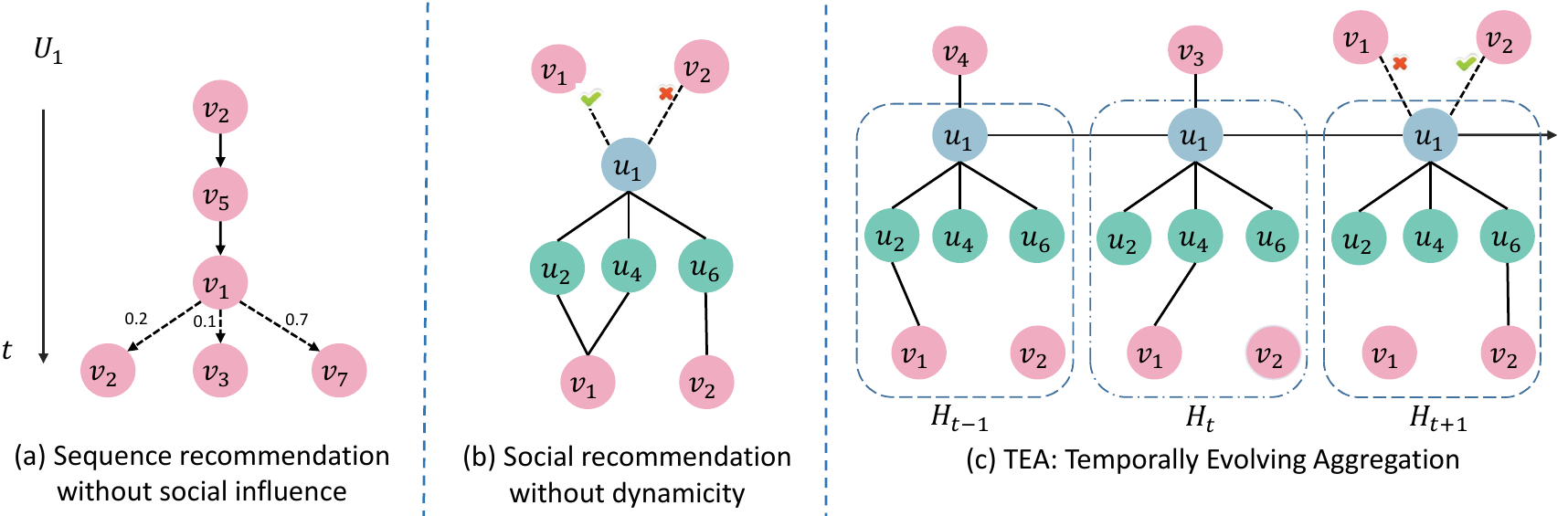}
	\caption{Illustration of a toy sequential recommendation example, where the blue node is the target user, the green nodes are the neighbors of the target user in the social network, and the pink nodes denote the items. The black dashed lines denote the anticipated recommended results between items and target user $u_1$. (a)-(b) Existing methods, which rarely consider the temporally dependent dynamics of the heterogeneous graph, lead to inaccurate prediction results. (c) On the contrary, the proposed \textbf{TEA} method simultaneously aggregates the historical user behavior sequence and the dynamic heterogeneous graph, thus resulting in more accurate predictions than existing methods.}
	\label{fig:motivation}
\end{figure*}

The sequential recommendation system is achieving more and more attention because of its practicality and effectiveness \cite{eskandanian2019modeling, he2017translation, tang2013exploiting, tang2016recommendation}. \textcolor{black}{In a sequential recommendation system,
the users access different items at different time stamps and frequently interact with each other.} The difficulties of sequential recommendation mainly come from two aspects:
% come from two kinds of sociological conclusion drawn from the observation of the crowd: 
\textcolor{black}{the temporal dependency of historical behaviors and the nonstationarity of users} \textcolor{black}{The temporal dependency of historical behaviors means that the decision of a user is influenced by the historical behaviors.
And the nonstationarity of users means that the decision of a user is influenced by the social relationship with the neighbors (\textcolor{black}{i.e. adjacent nodes of a user in the social networks}) and the user-item interactions of the neighbors.}
% The herd behaviors make people feel disposed to choose popular items, 
% and the homophily in the social networks enables recommendation algorithms to mine latent intents of users. 
Therefore, one important challenge is how to effectively leverage the historical behaviors and the social relationship between users.
% the user-item bipartite graphs and the social networks.

Focusing on the above challenge, numerous sequential recommendation algorithms have been proposed in recent years, which leverage the user behavior sequence and employ the Markov Chains to model the transition of items. Eskandanian et al. \cite{eskandanian2019modeling} mine the user preference and identify change-points
in the sequence of user interactions by using Hidden Markov Model (as shown in Figure \ref{fig:motivation} (a)). 
He et al. \cite{he2017translation} model the personalized sequential behavior by using the personalized translation vectors and the previous item embedding to predict the next item. \textcolor{black}{These transition-based methods assume that the users are independent of each other, which ignores the influence between the users. Considering that the behavior of a user is easily affected by the neighbors, ignoring the dependence between users will suffer from limited performance in the sequential recommendation.} 
% Hence, Straightforwardly employing the assumption will result in the suboptimal performance. }As a summary, the existing methods rarely consider the dependencies across the temporally evolving heterogeneous graphs and the user behavior sequence simultaneously. 

Another kind of recommendation algorithm \cite{fan2019graph, ma2011recommender, tang2016recommendation,jamali2010matrix} focuses on analyzing the social relationships between users and user-item interactions in a static user-item graph.
% , which consider the static network information \cite{fan2019graph, ma2011recommender, tang2016recommendation,jamali2010matrix}, have been proposed in recent years. One line of researches focuses on analyzing the user-item relationships and incorporating the mechanism of trust propagation. 
The typical methods include the traditional Collaborative Filtering (CF) methods \cite{hu2008collaborative, koren2008factorization, rendle2009bpr}, the deep learning enhanced approaches \cite{liu2020deep, wang2019multi, deng2019deepcf, covington2016deep}, the recently developed graph neural networks based methods \cite{battaglia2018relational}, as well as the social-network-based methods \cite{fan2019graph,song2019session,yu2020enhance}. These methods reveal that both the interactions among users in social networks and the user-item bipartite graphs are beneficial to the performance of the recommendation system. 
However, almost all the aforementioned methods assume the heterogeneous graphs of the users and the items are static, which ignores the dynamical influence \textcolor{black}{of the temporal interaction between items and social networks}, and further results in the suboptimal performance of recommendation systems. Take Figure \ref{fig:motivation} (b) for a toy example. The existing methods without considering the dynamic user-item heterogeneous graph might recommend the $v_1$ in preference to $v_2$ since more friends of user $U_1$ choose $v_1$. 

% Another line of recommendation methods is the sequence recommendation, which aims to learn the dynamic of user behavior sequence and utilize the Markov Chains to model the transition of items. Eskandanian et al. \cite{eskandanian2019modeling} mine the user preference and identify change-points
% in the sequence of user interactions by using Hidden Markov Model (as shown in Figure \ref{fig:motivation} (b)). 
% He et al. \cite{he2017translation} model the personalized sequential behavior by using the personalized translation vectors and the previous item embedding to predict the next item. However, these transition based methods assume that the users are independent, they usually ignores the temporally evolving interaction between items and the social networks resulting in the loss of helpful information about the communities of users. 
% As a summary, the existing methods rarely consider the dependencies across the temporally evolving heterogeneous graphs and the user behavior sequence simultaneously. 
% which cannot apply much rich information from the dependencies among items and the dynamic social influences for the social recommendation. 

Thus, it is essential to devise a unified framework to take advantage of both the historical behaviors of a user and dynamic interactions between the neighbors and items.  Figure \ref{fig:motivation} (c) illustrates our main idea that models the temporally user-item heterogeneous graphs and generates a more accurate prediction.
In the figure, the decision of whether a user $u_i$ will choose a given item $v_{t+1}$ is controlled by two important factors: (1) the historical interactions between him or her and items; (2) the temporal dynamic heterogeneous graph, including the interactions between the neighbors and items. Hence, the goal of the proposed method is to estimate the conditional probability of $v_{t+1}$ given a user $u_i$, the historical accessed item sequence $v_{1:t}$, as well as the heterogeneous graph sequence $H_{1:t+1}$, which can be formulated as $P(v_{t+1}|u_i, \mathcal{H}_{1:t+1}; v_{1:t})$.
% in which $\mathcal{H}_{1:t+1}$ is the temporally evolving heterogeneous graph and $v_{1:t}$ is the historical interactions sequence of $u_i$ with $t$ time steps. 

Based on the above idea, we propose the \textbf{T}emporally \textbf{E}volving \textbf{A}ggregation (\textbf{TEA} in short) framework for sequential recommendation by aggregating the user behavior sequence as well as the dynamic user-item heterogeneous graph. Inspired by the sequence labeling in natural language processing \cite{panchendrarajan2018bidirectional, hao2021semi} to model the joint probability distribution), we adopt CRF to model the item decision sequence and estimate $P(v_{t+1}|u_i, \mathcal{H}_{1:t+1}; v_{1:t})$. In order to alleviate the issue of the large item space, we use the pseudo likelihood method to approximate the aforementioned conditional probability. 
% Moreover, we borrow the idea of Word2Vec and finally drive a tractable and effective objective function.
\textcolor{black}{By doing this, the training procedure can be performed by estimating the unary score and transition score in CRF, which are implemented by our designed modules.}
{Technically, we design a \textit{Time-Restricted User Behavior Sequence Aggregation Module} to estimate the transition score of \text{CRF}, and a \textit{Temporal Dynamic Heterogeneous Graphs Aggregation Module }to estimate the Unary Scores of \text{CRF}.}
% Technically, we employ the personalized self-attention mechanism over historical user behavior sequences to estimate the item transition score. Moreover, we model the evolving social influence dependencies by aggregating the dynamic heterogeneous graph which includes the social networks and the dynamic user-item interactions. 
We further provide two different practical implementations based on the proposed framework. Extensive experimental studies demonstrate that our \textbf{TEA} framework outperforms the state-of-the-art recommendation methods on two published datasets and one real-world WeChat official accounts dataset.

\textcolor{black}{The contributions of this paper can be summarized as follows:
\begin{itemize}
    \item We formalize the sequential recommendation problem as a sequential decision problem that coincides with the CRF. To our best knowledge, this is the first attempt to study sequential recommendation based on CRF
    \item We improve the conventional CRF and propose a unified framework that simultaneously aggregates the historical user behaviors and the dynamic interactions between users. Moreover, we also provide two practice implementations. 
    \item we compare the proposed TEA framework with the 14 compared method on three real-world datasets and achieve state-of-the-art performance.
\end{itemize}}

% Based on the above formalization, we propose the \textbf{T}emporally \textbf{E}volving \textbf{A}ggregation (\textbf{TEA} in short) framework for social recommendation by aggregating the user behavior sequence and the dynamic social influence simultaneously. Inspired by the sequence labeling in natural language processing \cite{panchendrarajan2018bidirectional}, which uses conditional random field (CRF) to jointly model the label decision, we adopt CRF to model the items decision sequence and derive a tractable solution to estimate $P(v_{t+1}|u_i, \mathcal{H}_{1:t+1}; v_{1:t})$. In detail, we employ the personalized self-attention mechanism over historical user behavior sequences to estimate the item transition score. Moreover, we model the evolving social influence dependencies by aggregating the dynamic heterogeneous graph which includes the social networks and the dynamic user-item interactions. We further provide two different practical implementations based on the proposed framework. Extensive experimental studies demonstrate that our \textbf{TEA} framework outperforms the state-of-the-art social recommendation methods on two published datasets and one real-world WeChat official accounts dataset.

The remainder of this paper is organized as follows. In Section \ref{sec:related}, we review related researches into recommendation systems, including social recommendation and sequential recommendation. In Section \ref{sec:model}, we define the problem of sequential recommendation under the dynamical heterogeneous graph and further derive the objective function based on the conditional random field. In Section \ref{implementation}, we provide the implementation details of the proposed \textbf{TEA} model. We further analyze the connection to existing methods in Section \ref{connection}. And then, we present our experimental results based on two standard benchmarks and one real-world dataset in Section \ref{sec:exp}. Finally, we give our conclusion of the proposed method.

\section{Related Works} \label{sec:related}
In this section, we discuss the existing techniques on social recommendation and sequential recommendation 
\textcolor{black}{and the works about heterogeneous graph learning.}

In order to effectively mine the deep demands of users, researchers set their sights on social relations, hence social recommendation has received more and more attention. One of the most important methods is Matrix Factorization (MF) \cite{mnih2007probabilistic,baltrunas2011matrix,he2016fast}. Based on the traditional matrix factorization methods, Hao et al. \cite{ma2008sorec} proposed a co-factorization method, which shares a common latent user-feature matrix factorized by both ratings and social relations. With the development of deep learning methods, He et al. \cite{he2017neural} propose NeuMF by replacing the inner product with a neural architecture that can learn an arbitrary function from data. 
Fan et al. \cite{fan2018deep} propose a deep neural network-based model to learn non-linear features of each user from social relations and to integrate them into probabilistic matrix factorization for the social recommendation.
Deng et al. \cite{deng2016deep} propose a two-phase recommendation process to utilize deep learning to calculate the impact of community effect from the interests of users' trusted friends for recommendations.

Recently, graph neural networks (GNNs) \cite{battaglia2018relational,kipf2016semi} are widely used to aggregate node information and topological structure from social networks, hence GNNs are employed to address the social recommendation problem. In order to well aggregate the heterogeneous information, Fan et al. \cite{fan2019graph} propose the GraphRec for the social recommendation. Fu et al. \cite{fu2020magnn} leverage the metapaths \cite{shi2018heterogeneous} to obtain the heterogeneous graph embedding. Considering that the influences in the social network may be context-dependent, Song et al. \cite{song2019session} address the session-based social recommendation by using a dynamic-graph-attention neural network architecture. 
However, the aforementioned methods rarely consider the fact that different friends in social networks choose different items. In this work, considering the fact that social influence and user behaviors are time-dependent, the proposed \textbf{TEA} method focuses on aggregating the temporally evolving social influence and the user behavior sequence.

Since users usually access the items in chronological order, the users are likely to choose the items that are closely relevant to those they just accessed. Many works on sequential recommendation follow this assumption. Aiming to model the item-item transition probabilities, some traditional works borrow the idea of the Markov chain. Rendle et al. \cite{rendle2010factorizing} bridge the Matrix Factorization (MF) and Markov Chains (MC). He et al. \cite{he2017translation} propose TransRec to model such third-order relationships \textcolor{black}{(e.g. the relationships among a user, the previously accessed item and the next item)} for large-scale sequential prediction. Motivated by the advantages of sequence learning in natural language processing, many neural network-based methods are proposed to learn the sequential dynamics. Tang et al. \cite{tang2018personalized} leverage convolutional neural networks to encode the sequences into the embeddings. Other works \cite{hidasi2018recurrent, quadrana2017personalizing} leverage recurrent neural networks and their variants to model the sequences of items. Kang et al. \cite{kang2018self} further leverage attention-mechanism and propose the SASRec to balance the goal of MC-based methods and RNNs based methods. Moreover, Sun et al. \cite{sun2019bert4rec} argue that such left-to-right unidirectional models are sub-optimal. So they propose BERT4Rec, which employs deep bidirectional self-attention to model user behavior sequences.
% In this work, we borrow the idea of the Conditional Random Field (CRF) in named entities recognition and use CRF to model the user behavior dependencies.
\textcolor{black}{In this paper, the proposed \textbf{TEA} leverage the Conditional Random Field (CRF) to model the translation of items, which calculates the transition score and the unary score by respectively aggregating the user behavior sequence information as well as the dynamic user-item heterogeneous graph.}

\textcolor{black}{Our works are also related to the link prediction task of the heterogeneous graph since the recommending an item to a user can be considered as prediction a link between a user and an item in the bipartite graph. Schlichtkrull et.al \cite{schlichtkrull2018modeling} first introduce the relational graph convolutional networks (R-GCNs) to model the relation data and address the link prediction task in the heterogeneous graph. Recently, other researchers \cite{wang2019heterogeneous, hu2020heterogeneous} leverage the powerful attention mechanisms to select the key nodes for message-passing and further model the heterogeneous graph data. Considering that the heterogeneous graph usually contains many schema, Manchanda et.al \cite{manchanda2021schema} perform message passing to incorporate information of neighbors multiple hops away by taking advantage of the schemas of the heterogeneous graphs. } 
% Recently, disentangled representation \cite{bengio2013representation} is receiving more and more attention and applied on various fields\cite{cai2019learning,hao2021semi,ma2019learning,ma2019disentangled}. With the help of the spirit of the disentanglement, recommendation systems can enhance the robustness and the Interpretability. Ma et.al \cite{ma2019learning} use the variational inference to disentangle the high-level factors as well as the low-level factors to enhance the performance of recommendation system. Hu et.al \cite{hu2020graph} aim to disentangle a user’s latent preference factors which cause her clicks on different news, so they raise the graph neural news recommendation model with unsupervised preference disentanglement to encode high-order relationships into the disentangled representation. Wang et.al \cite{wang2020disentangled} also consider the user-item relationships at the finer granularity of user intents, so they propose the disentangled graph collaborative filtering, to disentangle these factors and yield disentangled representations. Wang et.al \cite{wang2020disenhan} argue that the existing heterogeneous graph neural networks neglect entanglement of the latent factors stemming from different aspects and the meta
% paths overlook the rich information of paths. Therefore, the propose the disentangled heterogeneous graph attention network and address the top-N recommendation by disentangling the user/item representations from different aspects in a heterogeneous information network.

\section{Model}\label{sec:model}
\newcommand{\tabincell}[2]{\begin{tabular}{@{}#1@{}}#2\end{tabular}}  
\begin{table}
\caption{Notation and Descriptions.}
	\centering
	\begin{tabular}{c|c}
		\hline
		\small{Notations}  & Descriptions \\
		\hline
		$U,V$            & User and item set. \\
		\hline
		$m,n$            & The size of user set and item set. \\
		\hline
		$\mathcal{G}^b, \mathcal{G}^b_t$  & \tabincell{c}{The bipartite graph only includes the user-item \\ interaction and that at the $t$-th timestamp.} \\
		\hline
		$\mathcal{E}^b, \mathcal{E}^b_t$  & \tabincell{c}{The edges set of bipartite graph \\ and that at the $t$-th  time-step.} \\
		\hline
		$\mathcal{G}^s$  & The social networks. \\
		\hline
		$\mathcal{E}^s$  & The edges among users in social network\\
		\hline
		$\mathbf{H}_t$  & \tabincell{c}{The heterogeneous graph that includes the \\ social network $\mathcal{G}^s$ and the bipartite graph $\mathcal{G}^b_t$ \\ at $t$-th time-step.} \\ 
		\hline
		$\mathbf{p}_i$   & The embedding of user $u_i$. \\
		\hline
		$\mathbf{q}_j$   & The embedding of item $v_j$. \\
		\hline
		$\mathbf{k}_j$   & The embedding of the $j$-th position in item sequences.\\
		\hline
		$\mathbf{W}, \mathbf{b}$ & Weights and biases in neural networks. \\
		\hline
		$d$              & The dimension number of representation. \\
		\hline
		$\Theta_f$       & The parameters of unary scores function. \\
		\hline
		$\Theta_g$       & The parameters of transition scores function. \\
		\hline
		$\oplus$         & The concatenation operator of any two vectors. \\
		\hline
		$\bm{x}$         & The observed item sequence.\\
		\hline
		$\bm{y}$         & The label item sequence of $\bm{x}$ \\
		\hline
		$\mathcal{N}(u_i)$    & The 1st-order neighbourhood of user $u_i$ \\
		\hline
		$\mathcal{I}_t({u_i})$    & The accessed items of user $u_i$ at $t$-th time-step. \\
		\hline
		$\tau$    & \tabincell{c}{Time window for selecting the walks in the duration \\ of [t-$\tau$, t+$\tau$].} \\
		\hline
		$d$    & The dimension of user and item embedding. \\
		\hline
	\end{tabular}
	\label{tab:notation}
\end{table}
% \vspace{3ex}
In this section, we begin with the problem definition of sequential recommendation. Then we derive the unified objective function under conditional probability $P(v_{t+1}|u_i, \mathcal{H}_{1:t+1}; v_{1:t})$.
% Finally, we provide the implementation details of the \textbf{TEA} framework.

\subsection{Problem Definition}
Let $U=\{u_1, u_2, \cdots, u_n\}$ and $V=\{v_1, v_2, \cdots, v_m\}$ denote the sets of users and items respectively, in which $n$ is the number of users and $m$ is the number of items. For user-item interactions, we let $\mathcal{G}^b=\{U \cup V, \mathcal{E}^b\}$ be the user-item bipartite graph with edges $(u_i, v_j) \in \mathcal{E}^b$. As for user-user relations, we let $\mathcal{G}^s=\{U, \mathcal{E}^s\}$ be the social graph with edges $(u_i, u_j) \in \mathcal{E}^s$. If we combine the bipartite graph and the social graph, we obtain the following heterogeneous graph $\mathbf{H}=\{U \cup V, \mathcal{E}^b \cup \mathcal{E}^s\}$. Let $v_{1:t}$ be the user behaviors sequence for $u_i$. Since we consider the temporal evolving social influence, we let $\mathcal{H}_{t+1}=\{\mathbf{H}_1, \mathbf{H}_2, \cdots, \mathbf{H}_{t+1}\}$ be the heterogeneous graph sequence, where $\mathbf{H}_t=\{U \cup V, \mathcal{E}^b_t \cup \mathcal{E}^s\}$ and $\mathcal{E}^b_t$ is the user-item interactions in $t$-th time-step. For user $u_i$, given the behavior sequence $v_{1:t}$ and the heterogeneous graph sequence $\mathcal{H}_{t+1}$ as well as the item $v_{t+1}$, our goal is to estimate the conditional probability of $P(v_{t+1}|v_{1:t},u_i,\mathcal{H}_{t+1})$. The mathematical notation and the corresponding descriptions are summarized in Table \ref{tab:notation}. %$P(v_{t+1}|u_i, \mathcal{H}_{t+1}, S^{u_i}_t)$. 

% 1. user item，user-item组成二部图，user-user社交网络图，合起来是一个异构图
% 2. 两个序列，item行为序列，动态图序列（N阶邻居+邻居的购买商品行为构成的异构图）
% 3. 推荐是一个user-item矩阵填充问题，给定两个序列，填充矩阵
% 4. 符号表

%\subsection{Preliminary Knowledge}
\subsection{Methodology}
%CRF是什么，有什么优点，为什么用？
%怎么用CRF，直接摆出损失函数
We begin with the traditional Conditional Random Field (CRF), which is a probabilistic graphical model widely used in sequence labeling \cite{panchendrarajan2018bidirectional}. CRF has shown to be very effective since it can jointly model the label decision by capturing the dependencies across adjacent labels. Considering the general definition of CRF, let $\bm{x}=\{x_1, \cdots,x_t,\cdots, x_T\}$ and $\bm{y}=\{y_1, \cdots, y_t, \cdots, y_T\}$ denote the observed sequence and its corresponding labels respectively. Formally, the conditional distribution $p(\bm{y}|\bm{x})$ of Linear Chain CRF\cite{ma2016end} is given by:
\begin{equation}
\label{equ:crf}
\begin{split}
p(\bm{y}|\bm{x}) &= \frac{1}{Z(\bm{x})}\exp(\sum_{t=1}^{T}f(x_t,y_t;\Theta_f) + \sum_{t=1}^{T-1}g(y_t, y_{t-1};\Theta_g)), \\
%E(\mathbf{y}|\mathbf{x}) &= \sum_{t=1}^{T}f(x_t,y_t) + \sum_{t=1}^{T-1}g(y_t, y_{t-1}).
Z(\bm{x}) &= \sum_{\bm{y'}}\exp(\sum_{t=1}^{T}f(x_t,y'_t;\Theta_f) + \sum_{t=1}^{T-1}g(y'_t, y'_{t-1};\Theta_g)),
\end{split}
\end{equation}
in which $\Theta_f$ and $\Theta_g$ are the trainable parameters.

There are three important components in the above CRF model: the partition function $Z(\bm{x})$, the unary scores function $f(x_t,y_t)$ and the transition scores function $g(y_t, y_{t-1})$. The partition function $Z(\bm{x})$ is a normalization factor in order to obtain a probability. The unary scores function $f(x_t,y_t)$ is used to estimate the probability of $y_t$ given the observed $x_t$. And the transition scores function $g(y_t, y_{t-1})$ is used to estimate the probability of $y_t$ given $t_{t-1}$.

The three components framework provides us a unified solution to aggregate both the historical behaviors of users and the dynamic social influence from the social networks. Following the formulation of CRF, the purpose of our model is to estimate the conditional distribution as follows:
\begin{equation}
\label{equ:social_crf}
\begin{split}
P(v_{1:t+1}|u_i, \mathcal{H}_{t+1}) &= \\
\frac{1}{Z(\mathcal{H}_{t+1}, u_i)}\exp(&\sum_{t=1}^{T}f(\mathbf{H}_{t+1}, u_i, v_{t+1};\Theta_f)\\&+\sum_{t=1}^{T-1}g(v_{t+1}, v_{t};\Theta_g)), \\
Z(\mathcal{H},u_i)=\sum_{S'^{u_i}_{t+1}}\exp(&\sum_{t=1}^{T}f(\mathbf{H}_{t+1}, u_i, v_{t+1};\Theta_f)\\&+\sum_{t=1}^{T-1}g(v_{t+1}, v_{t};\Theta_g)),
\end{split}	
\end{equation}
in which $f(\mathbf{H}_{t+1}, u_i, v_{t+1};\Theta_f)$ denotes the aggregation of temporal evolving social influence, $g(v_{t+1}, v_{t};\Theta_g)$ denotes the aggregation of user behaviors. In specific, $f(\mathbf{H}_{t+1}, u_i, v_{t+1};\Theta_f)$ describes the relationship between the dynamic heterogeneous graph $\mathbf{H}_{t+1}$ and the available item $v_{t+1}$ and $g(v_{t+1}, v_{t};\Theta_g)$ models the dependency between the available item $v_{t+1}$ and the user behavior sequence.

However, it is almost impossible to calculate $Z(\mathcal{H}, u_i)$ since the sequence length is too large. In order to address this issue, we employ the pseudo likelihood method as an effective approximation method \cite{besag1975statistical,ma2018cgnf}, and further derive the following estimation of the conditional probability:
\begin{equation}
\label{equ:social_crf_app}
\begin{split}
P(v_{1:t+1}|u_i,\mathcal{H}_{t+1}) \approx PL(v_{1:t+1}|u_i,\mathcal{H}_{t+1}) =\\ \prod \limits_{t} P(v_{t+1}|v_{1:t},u_i,\mathcal{H}_{t+1}).
\end{split}
\end{equation}

Combining Equation (\ref{equ:social_crf}) and Equation (\ref{equ:social_crf_app}), we further derive the following estimation of the conditional probability $P(v_{t+1}|v_{1:t},u_i,\mathcal{H}_{t+1})$:
\begin{equation}
\begin{split}
P(v_{t+1}|v_{1:t}&,u_i,\mathcal{H}_{t+1})=\\
&\frac{\exp(f(\mathbf{H}_{t+1}, u_i, v_{t+1};\Theta_f)+g(v_{t+1}, v_{1:t};\Theta_g))}
{\sum_{v_j\in V}\exp(f(\mathbf{H}_{t+1}, u_i, v_j;\Theta_f)+g(v_j, v_{1:t};\Theta_g))}.
\end{split}
\end{equation}

Finally, we can obtain the objective function of our proposed model as follows:
\begin{equation}
\mathcal{L}_{crf} = \frac{1}{n}\sum_{i=1}^{n} \sum_{t=1}^{T}\log P(v_{t+1}|v_{1:t},u_i,\mathcal{H}_{t+1}).
\end{equation}

The aforementioned objective function is usually impractical because the size of the item set is very large and the computation cost is unaffordable. 
% There are many methods to address this problem \cite{mikolov2013distributed}, like  Hierarchical Softmax and negative sampling. In this paper, by employing the negative sampling strategy similar to Word2Vec, we finally 
Inspired by \cite{mikolov2013distributed}, we employ the negative sampling strategy to obtain the tractable unified objective function of sequential recommendation as follows:
\begin{equation}
\label{equ:final_loss}
\begin{split}
\mathcal{L}_{crf} =&\frac{1}{n}\sum_{i=1}^{n}\sum_{t=1}^{T-1} \log\sigma(f(\mathbf{H}_{t+1}, u_i, v_{t+1};\Theta_f) \\&+ g(v_{t+1}, v_{1:t};\Theta_g)) +\\
& \sum_{k=1}^{n_s}[\log\sigma(-f(\mathbf{H}_{t+1}, u_i, v_{k};\Theta_f) \\&- g(v_{k}, v_{1:t};\Theta_g))],
\end{split}
\end{equation}
where $\sigma$ is the sigmoid activation function and $v_k$ is the negative item uniformly sampled from the whole item set $V$.

% We can find that the objective function shown in Equation (\ref{equ:final_loss}) .
The objective function enjoys an \textcolor{black}{appealing} physical meaning.
It provides the insights into how to design the model for sequential recommendation:  $f(\mathbf{H}_{t+1}, u_i, v_{t+1};\Theta_f)$ models the information of temporal evolving heterogeneous graph in the forms of the unary energy function; meanwhile  $g(v_{t+1}, v_{1:t};\Theta_g)$ not only models the alternative item $v_t$ but also the user behavior sequence in the form of the pairwise energy function. 
\textcolor{black}{Based on the aforementioned objective function, we further illustrate the training process of the model shown in Algorithm 1.}
\begin{algorithm}[htb]
	\renewcommand{\algorithmicrequire}{\textbf{Input:}}
	\renewcommand{\algorithmicensure}{\textbf{Output:}}
	\caption{\textcolor{black}{TEA training algorithm}}
	\label{alg1}
	\begin{algorithmic}[1]
		\REPEAT
        \STATE \textcolor{black}{Randomly select a batch of the users, then select the accessed items and the corresponding heterogeneous graph sequence;} 
	    \STATE \textcolor{black}{Forward propagation;} 
	    \STATE \textcolor{black}{Update $\Theta_f$ and $\Theta_g$ based on Equation (6); }
	   % \STATE Augment $g_s, g_t$ based on Equation \eqref{equ:noise_gen}
	   % \STATE Forward propagation; 
	   % \STATE Update $\phi_{f}, \phi_{o}, \theta_{o}$ based on Equation \eqref{equ:update2}; 
		\UNTIL \textcolor{black}{Max Iteration or Early Stopping.}
	\end{algorithmic}  
\end{algorithm}

\begin{figure*}[htbp]
	\includegraphics[width=\textwidth]{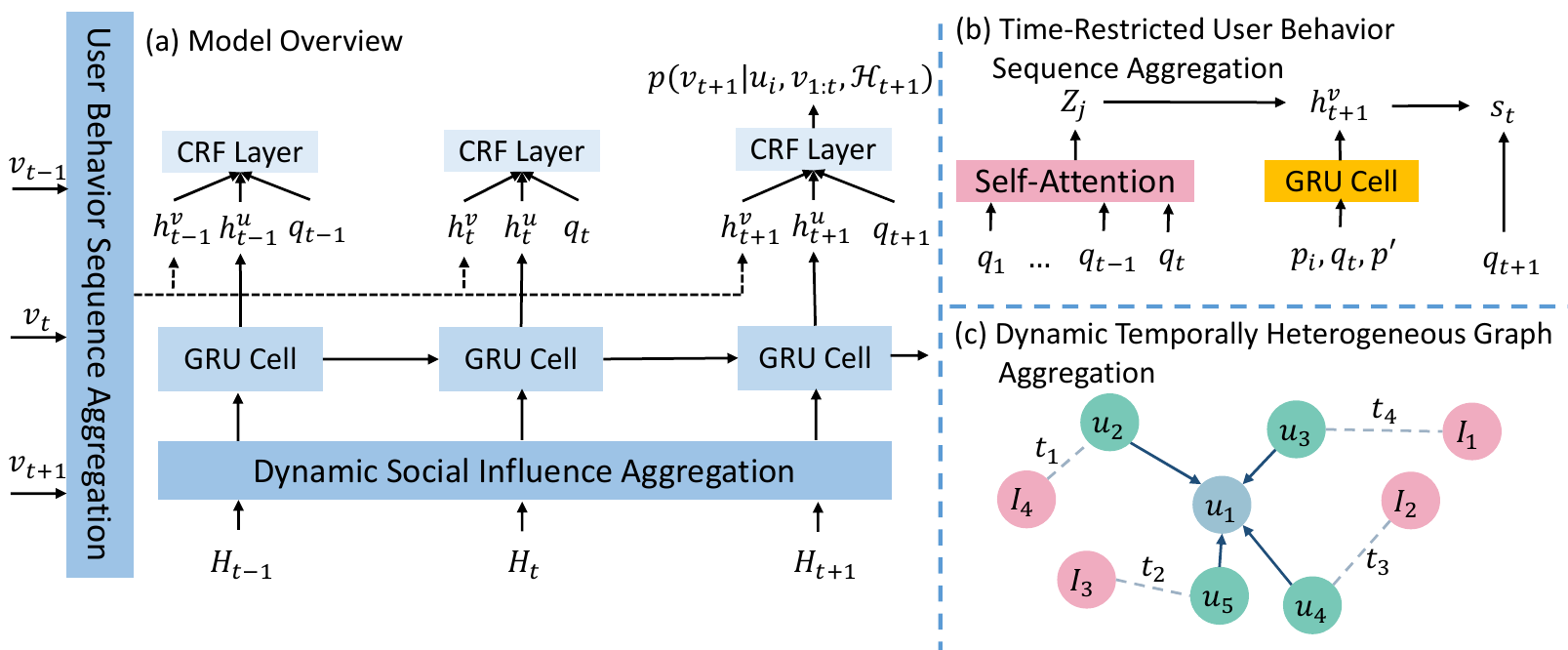}
	\caption{The framework of the temporally evolving aggregation model for the sequential recommendation. (a) The overview of the proposed model, the temporally dependent heterogeneous graphs aggregated representation $h_{t+1}^u$, the user behavior aggregated representation $h_{t+1}^v$ and the item embedding $\mathbf{q}_{t+1}$ are fed into the CRF layer and $P(v_{t+1}|u_i, \mathcal{H}_{t+1}; v_{1:t})$ is estimated. (b) The time-restricted user behavior sequence aggregation block is based on the user behavior sequence aggregation and the time-restricted aggregation. Note that the GRU used in this module is different from that in (a). (c) The dynamic temporally heterogeneous graph aggregation block, which is based on the bipartite graph aggregation and social network aggregation, takes $\mathcal{H}_t$ as input, the arrows denote the message passing direction.}
	\label{fig:model}
\end{figure*}
\section{Implementation of Temporally Evolving Aggregation Framework}\label{implementation}

% \vspace{-2ex}
In this section, we provide the implementation details of the proposed temporally evolving aggregation model. As illustrated in Figure \ref{fig:model}(a), our implementation takes both the aggregation of user behavior sequences and the aggregation of temporally dependent heterogeneous graphs into consideration and employs the GRU cells \cite{cho2014learning} and CRF layers to predict the final results. The details of the two aggregation modules are presented in Figure \ref{fig:model} (b) and Figure \ref{fig:model} (c) respectively. We will give detailed descriptions of these two aggregation modules in the following subsections. 

\subsection{Time-Restricted User Behavior Sequence Aggregation for the Transition Scores}
In this subsection, we will introduce the technical details of $g(v_{t+1}, v_{1:t};\Theta_g)$. Given user $u_i$ and the corresponding behavior sequence $v_{1:t}$, we aim to calculate the user-specific item transition score. 

\subsubsection{User Behavior Sequence Aggregation}
Considering that the future behavior of a user is not only influenced by the latest accessed items but also the items that the user has accessed before, the user behavior sequence aggregation block should consider both the transition between items and the long-term dependency of items. Inspired by the great success of the self-attention mechanism \cite{vaswani2017attention} in various tasks like machine translation, we propose an extension of the self-attention mechanism to model the personalized item transition and long-term dependency by simultaneously leveraging the item information and the position information. Formally, given the $j$-th candidate item, we calculate the weights of each historical item as follows:
\begin{equation}
\begin{split}
    a_{\tau j} = \text{softmax}(\frac{\mathbf{W}_{Q}(\mathbf{q}_j+\mathbf{k}_j) \left(\mathbf{W}_{K} (\mathbf{q}_\tau + \mathbf{k}_\tau) \right)^{\mathsf{T}}}{\sqrt{d}}), &\tau < j ,
\end{split}
\end{equation}
where $\mathbf{q}_j$ is the embedding of item $v_j$, $\mathbf{k}_j$ is the position embedding at $j$-th position of the input sequence, $\mathbf{W}_Q, \mathbf{W}_K$ are trainable projection parameters and $\sqrt{d}$ is the scaling factor, and $d$ is the dimension of the embedding. As a result, we can calculate the historical item aggregated representation as follows:
\begin{equation}
\mathbf{z}_j = \sum_{\tau=1}^{\tau=j-1} a_{\tau j}\mathbf{W}_V\left(\mathbf{q}_\tau + \mathbf{k}_\tau\right),
\end{equation}
in which $\mathbf{W}_V$ are trainable projection parameters. 

\subsubsection{Time-Restricted Aggregation}
Since the temporal interactions between users and items are very sparse, for the users that contain limited social relationships and items interactions, it is hard to obtain an ideal user embedding for the sparse social substructure, and it is also difficult to obtain a debiased item embedding. Therefore, it is a challenging task to well aggregate the information from the users to the items and vice verse. Fortunately, we find that the users that select the same items usually share the same interests and intent. Inspired by this intuition, we further proposed the time-restricted aggregation module.

First, we selected the walk with three nodes (e.g., USER-ITEM-USER) with the restriction of time window $\tau$. In detail, given the interaction $(u_i, v_t)$, we find the other users that select the same item in the time window of $[t-\tau, t+\tau]$, where $\tau$ is the window size. In our experimental implementation, we choose $\tau=60$ days. Therefore, we can collect the $\tau-$restricted walks for example $u_i-v_t-u'$. Sequentially, we employ another GRU to aggregate the information from the dense substructures to the sparse substructures, which can be formalized as follow:
\begin{equation}
    \mathbf{h}_{u_i}, \mathbf{h}_{v_t}, \mathbf{h}_{u'} = \text{GRU}(\mathbf{p}_i,\mathbf{q}_t,\mathbf{p}';\mathcal{W}_R).
\end{equation}
% \blue{bm font?}
\textcolor{black}{Note that we take the whole sequence $u_i-v_t-u'$ as input of a GRU layer instead of a GRU cell and $\mathbf{h}_{u_i}, \mathbf{h}_{v_t}, \mathbf{h}_{u'}$ are the output corresponding to $\bm{p}_i,\mathbf{q}_t$ and $\mathbf{p'}$ respectively. $\mathcal{W}_R$ are the trainable parameters.}
% in which we take the walk $u_i-v_t-u'$ as input and $\mathbf{h}_{u_i}, \mathbf{h}_{v_t}, \mathbf{h}_{u'}$ are the output of GRU of each timestamp; $\mathcal{W}_R$ are the trainable parameters.

\subsubsection{Calculate the Transition Scores}
In order to well perform the personalized user behavior sequence aggregation, we further add the user embedding $\mathbf{p}_i$ into the transformed item representation. Formally, we can calculate the transition score $s_t$ as follows:
\begin{equation}
\begin{split}
s_t &= \left(\mathbf{W}_g^{(3)}\left[\mathbf{h}_t^{v_j}\oplus \bm{h}_{u_i} \oplus  \bm{h}_{v_t}\oplus \mathbf{p}_i \right]\right)^\mathsf{T} \bm{q}_j, \\
\mathbf{h}_t^{v_j} &= \mathbf{p}_i + \mathbf{W}_g^{(2)}\left(\text{ReLU}(\mathbf{W}_g^{(1)}\mathbf{z}_j  + \mathbf{b}_g^{(1)}) \right) + \mathbf{b}_g^{(2)},
% s_t &= {\mathbf{h}_t^{v_j}}^\mathsf{T} \mathbf{q}_j,
\end{split}
\end{equation}
in which $\mathbf{W}_g^{(1)}, \mathbf{W}_g^{(2)}, \mathbf{b}_g^{(1)}, \mathbf{b}_g^{(2)}$ are the trainable parameters. For convenience, we let $\Theta_g=\{\mathbf{W}_Q, \mathbf{W}_K, \mathbf{W}_V, \mathbf{W}_g^{(1)}, \mathbf{W}_g^{(2)}, \mathbf{W}_g^{(3)},\mathbf{b}_g^{(1)}, \mathbf{b}_g^{(2)}, \mathbf{p}, \mathbf{q}, \mathbf{k},\bm{\omega}_R\}$ be the trainable parameters of $g(v_{t+1}, v_{1:t};\Theta_g)$.

\subsection{Dynamic Temporal Heterogeneous Graphs Aggregation for the Unary Scores}
In this part, we will introduce the details of the dynamic temporally heterogeneous graphs aggregation $f(\mathbf{H}_{t+1}, u_i, v_{t+1};\Theta_f)$, which is used to calculate the unary scores. The dynamic temporally heterogeneous graphs aggregation contains the bipartite graph aggregation and the social network aggregation. 

\subsubsection{Bipartite Graph Aggregation} In this part, we aim to obtain the aggregated of the bipartite graph at $t$-th time-step. Given user $u_i$ and the heterogeneous graph sequence $\mathcal{H}_{t+1}$, we first obtain the user-specific representation $\hat{\mathbf{h}_{t}}$ of $\mathcal{H}_t$. Specifically, we employ two different aggregated strategies and raise two variants of the proposed method: the GraphSAGE \cite{hamilton2017inductive} based method (named TEA-S) and the graph attention networks \cite{velivckovic2017graph} based method (named TEA-A). More experimental details will be introduced in the next section.

% \subsubsection{Two Variants}
As for the TEA-S variation, we can obtain the user-specific representation $\hat{\mathbf{h}_{t}}$ as follows: 
\begin{equation}
\hat{\mathbf{h}_{t}} = \text{ReLU} \left({\mathbf{W}_{A}} \operatorname{MEAN}\left(\mathbf{q}_{k}, \forall k \in \mathcal{I}_t(\mathcal{N}(u_i))\right)\right) ,
\end{equation}
where $\mathbf{W}_{A}$ are the trainable parameters and $\mathcal{I}_t(\mathcal{N}(u_i))$ denotes the items interacted by $u_i$'s neighbors at between $t$-th and $t+1$-th time-step; and $\operatorname{MEAN}$ denotes the average pooling operation.

As for the TEA-A variation, we aggregate the item information to the user with the help of the graph attention mechanism, which can be formulated as: 
\begin{equation}
\begin{split}
&\qquad\qquad\hat{\mathbf{h}_{t}} = \text{ReLU} \left(\sum_{j \in \mathcal{I}_t(\mathcal{N}(u_i)) } \alpha_{i j}\mathbf{q}_{j}\right),\\
\end{split}
\end{equation}
{where $\alpha_{ij}$ is the weight of user $u_i$ and item $v_j$ and is defined as}
\begin{equation}
\small
\begin{split}
\alpha_{i j} = &\frac{\exp \left(\operatorname{LeakyReLU}\left({\mathbf{w}_{A}}^{\mathsf{T}}\left[ {\mathbf{W}_{A}\mathbf{q}_{t}} \oplus  \mathbf{W}_{A}\mathbf{q}_{j}\right]\right)\right)}{\sum_{k \in \mathcal{I}_t(\mathcal{N}(u_i)) } \exp \left(\operatorname{LeakyReLU}\left({\mathbf{w}_{A}}^{\mathsf{T}}\left[ \mathbf{W}_{A}\mathbf{q}_{t} \oplus  \mathbf{W}_{A}\mathbf{q}_{k}\right]\right)\right)}
,
\end{split}
\end{equation}
% \green{where \textcolor{black}{$\alpha_{ij}$ is the weight of user $u_i$ and item $v_j$}; }
in which $\mathbf{q}_t$ is the embedding of the item interacted by $u_i$ at $t$-th time-step and $\oplus$ is the concatenation operation. $\mathbf{w}_{A}$ and $\mathbf{W}_{A}$ are trainable parameters. And $LeakyReLU$ is the leaky version of a rectified linear unit.

In order to model temporally dependent heterogeneous graphs propagation, we feed $\hat{\mathbf{h}_t}$ into the Gated Recurrent Unit \cite{cho2014learning}. The GRU cell operation at the $t$-th time-step can be formulated as: 
% \begin{equation}
%     \begin{array}{l}
%         z_{t} =\sigma\left(\hat{\mathbf{h}_{t}} \mathbf{W}_u_{z} + \mathbf{h}_{t-1} \mathbf{W}_{z}\right) \\
%         r_{t} =\sigma\left(\hat{\mathbf{h}_{t}} \mathbf{U}_{r} + \mathbf{h}_{t-1} \mathbf{W}_{r}\right) \\
%         \mathbf{h}_{t}^{\prime} =\tanh \left(\hat{\mathbf{h}_{t}} \mathbf{U}_{h}+\left(\mathbf{h}_{t-1} \circ r_t\right) \mathbf{W}_{h}\right) \\
%         \mathbf{h}_{t}=(1-z_t) \circ \mathbf{h}_{t}^{\prime}+z_t \circ \mathbf{h}_{t-1}
%     \end{array}
% \end{equation}
% in which $\mathbf{U}_{z}, \mathbf{W}_{z}, \mathbf{U}_{r}, \mathbf{W}_{r}, \mathbf{U}_{h}, \mathbf{W}_{h}$ are trainable parameters.
\begin{equation}
\mathbf{h}_t = \text{\textcolor{black}{GRUCell}}(\hat{\mathbf{h}_t}, \mathbf{h}_{t-1}; \mathcal{\bm{W}}_G),
\end{equation}
in which $\mathcal{\bm{W}}_G$ denotes all trainable parameters of the GRU cell. \\ 
\subsubsection{Social Network Aggregation}
To propagate the information of neighbors' interests, we further aggregate the information from the social network. For simplicity, we only formulate the GraphSAGE aggregation as follows: 
\begin{equation}
\mathbf{h}_{s} = \text{ReLU}\left(\mathbf{W}_{S} \operatorname{MEAN}\left( \mathbf{p}_{k}, \forall k \in \mathcal{N}(u_i) \right)\right),
\end{equation}
where $\mathbf{W}_{S}$ is the trainable parameters. 

% Finally, we fuse the time-dependent representation $\mathbf{h}_t$ and time-independent representation $\mathbf{h}_s$ into one vector and calculate the social influence score $s_f$, i.e., the output of unary scores function $f(\cdot)$. It is formulated as: 
% \begin{equation}
% \begin{split}
% \mathbf{h}_t^{u_i} &= \text{ReLU}( [\mathbf{h}_{t} \oplus \mathbf{h}_{s} ] \mathbf{W}_f^{(1)} + \mathbf{b}_f^{(1)}) \mathbf{W}_f^{(2)} + \mathbf{b}_f^{(2)},\\
% s_f &= {\mathbf{h}_t^{u_i}}^\mathsf{T} \mathbf{q}_j,
% \end{split}
% \end{equation}in which $\mathbf{W}_f^{(1)}, \mathbf{W}_f^{(2)}, \mathbf{b}_f^{(1)}, \mathbf{b}_f^{(2)}$ are trainable parameters. 

\subsubsection{Calculate the Uunary Scores}
Based on the aforementioned aggregation, we fuse the time-dependent representation $\mathbf{h}_t$ and time-independent representation $\mathbf{h}_s$ into one vector and calculate the social influence score $s_f$, i.e., the output of unary scores function $f(\cdot)$. It is formulated as: 
\begin{equation}
\begin{split}
s_f &= {\mathbf{h}_t^{u_i}}^\mathsf{T} \mathbf{q}_j,\\
\mathbf{h}_t^{u_i} = \mathbf{W}_f^{(2)}\text{ReLU}&(\mathbf{W}_f^{(1)} [\mathbf{h}_{t} \oplus \mathbf{h}_{s} ]  + \mathbf{b}_f^{(1)}) + \mathbf{b}_f^{(2)},
\end{split}
\end{equation}in which $\mathbf{W}_f^{(1)}, \mathbf{W}_f^{(2)}, \mathbf{b}_f^{(1)}, \mathbf{b}_f^{(2)}$ are trainable parameters. In summary, we let $\Theta_f=\{ \mathbf{W}_{A}, \mathbf{W}_{S}, \bm{\omega}_{G},\mathbf{W}_f^{(1)}, \mathbf{W}_f^{(2)}, \mathbf{b}_f^{(1)}, \mathbf{b}_f^{(2)}, \mathbf{p}, \mathbf{q}\}$ be the trainable parameters of $f(\mathbf{H}_{t+1}, u_i, v_{t+1};\Theta_f)$.

\subsection{Model Summarization}

%Finally, our model makes prediction by summing up the transition score $s_g$ and the social influence score $s_f$, which can be formulated as 

The total loss of our proposed model is summarized as follow:
\begin{equation}
\mathcal{L} = \mathcal{L}_{crf} + \gamma \mathcal{L}_{reg},
\end{equation}
where $\mathcal{L}_{reg}$ is the L2 normalization on all parameters and $\gamma$ is a trade-off hyper-parameter. 

Based on this objective function, our model is trained by the following procedure:
\begin{equation}
(\hat{\Theta_g},\hat{\Theta_f}) = \underset{\Theta_g, \Theta_f}{\arg \min }\mathcal{L}.
\end{equation}
All parameters are jointly optimized using the Adam\cite{kingma2014adam} algorithm. 

In the testing, we estimate the probability of $P(v_{t+1}|v_{1:t},u_i,\mathcal{H}_{t+1})$ as follows:
\begin{equation}
\begin{split}
P(&v_{t+1}|v_{1:t},u_i,\mathcal{H}_{t+1}) =\\&\sigma(f(\mathbf{H}_{t+1}, u_i, v_{t+1};\hat{\Theta_f})+g(v_{t+1}, v_{1:t};\hat{\Theta_g})).
\end{split}
\end{equation}

% In the testing, we estimate the probability of $P(v_t|v_{1:t-1},u_i, \mathbf{H})$ as follows:
% \begin{equation}
%     P(v_t|v_{1:t-1},u_i, \mathbf{H}) = \sigma(f(\mathcal{H}_t, u_i, v_t;\Theta_f) + g(v_t, v_{1:t-1};\Theta_g)).
% \end{equation}

\section{Connections to Existing Models}\label{connection}
We will discuss the connections to the existing transition-based sequential recommendation methods. Most of the existing works of transition-based sequential recommendation methods \cite{he2017translation,rendle2010factorizing} are based on Markov Chains. These methods mainly consider two important factors: (1) the interactions between users and items to capture the inherent intent of users, (2) the sequential dynamics between items to capture the relationships between items. Thus, we find that our method is more general and some of the existing works can be taken as special cases of ours. The detailed discussions for each work are as follows. 

Regarding the work FPMC \cite{rendle2010factorizing}, it simplifies the huge state space problem by introducing the basket of items and consequently ignores the sequence information of historical items in each basket. In the contrast, our method utilizes the historical item sequence by using the self-attention mechanism with position embedding and is more general than FPMC.

Regarding the work TransRec
\cite{he2017translation}, it models the personalized sequential behavior by using the personalized translation vectors and the previous item embedding to predict the next items but ignores the long-term dependencies since it only considers the relationships between any two items. Moreover, TransRec addresses the problem of the huge state space of items by introducing the subspace, while our method utilizes the negative sampling strategy. Thus, our method is more feasible and efficient to capture the dynamic social influence of the target users.

\section{Experiment}\label{sec:exp}

In this section, we experimentally evaluate the performance of our method on three datasets against the state-of-the-art compared methods. The preprocessed scripts and the source code can be found at \footnote{{https://github.com/DMIRLAB-Group/TEA}}.
\begin{table}[t]
	\caption{Statistics of the datasets.}
	\label{tab:stat}
	\centering
	\scalebox{1.0}{
		\begin{tabular}{p{2.0cm}p{1.5cm}p{1.5cm}p{1.5cm}}
			\toprule
			Dataset & \textbf{Epinions} & \textbf{Yelp} & \textbf{Wechat}\\
			\midrule
			\# users & 22,167 & 270,770 & 568,100 \\
			\# items & 296,278 & 184,134 & 242,702 \\
			\# interactions & 798,620 & 3,602,495 & 9,422,722 \\
			\# social links & 355,813 & 5,974,526 & 5,667,864 \\
			density & 0.0121\% & 0.0072\% & 0.0068\% \\
			social density & 0.0724\%	& 0.0081\% & 0.0018\% \\
			\bottomrule
		\end{tabular}
	}
\end{table}
% \vspace{-3ex}
\subsection{Datasets}
We evaluate our proposed TEA framework on two public datasets (Epinions and Yelp) and a large-scale private dataset (WeChat Official Accounts Dataset). 
The statistics of datasets are summarized in Table \ref{tab:stat}. The brief information of the datasets is as follows: 
\begin{itemize}
	\item {Epinions}\footnote{{http://www.trustlet.org/extended\_epinions.html}}: A benchmark dataset for the recommendation. In Epinions, a user can rate and give comments on items. Besides, a user can also select other users as their trusters, and we use the trust graphs \textcolor{black}{(which are composed of the trust relationships)} as the network information. 
	\item {Yelp}\footnote{{https://www.kaggle.com/yelp-dataset/yelp-dataset}}: An online review platform where users review local businesses (e.g., restaurants and shops). The user-item interactions and the social networks are extracted in the same way as Epinions.  
	\item {WeChat Official Accounts Dataset}: WeChat is a Chinese multi-purpose messaging, social media, and mobile payment application developed by Tencent. And WeChat official accounts dataset is one of the functions. On the WeChat Official Account platform, users can read and share articles. This dataset is constructed by user-article clicking records and user-user social networks on this platform. 
\end{itemize}

We preprocess the datasets following the approach in \cite{he2017translation}. \textcolor{black}{Specifically, for all these datasets, we follow the previous works \cite{kang2018self,sun2019bert4rec} and treat a rating or review as implicit feedback.} We further use the timestamps to determine the sequence order of actions. We discard users and items with fewer than 5 associated actions. In cases where star ratings are available, we take the item with a rating higher than 3 as users’ positive feedback. 
%For all these datasets, we assume the social relationship is undirected. For Epinions and Yelp, we convert explicit ratings$\geq 3$ on items into 1 as the implicit feedback. 

For data splitting, we employ the widely used leave-one-out evaluation \cite{rendle2009bpr, he2017neural}. We hold out the latest interaction of each user as the test set and select the second latest interaction as the validation set. The remaining data are used for training. 

\begin{table*}[htb]
	\caption{The performance evaluation of the compared methods on Epinions dataset. The values presented are averaged over 5 replicated with different random seeds.}
% 	\small
	\label{tab:epin}
	\centering
	\scalebox{1.1}{
	\begin{tabular}{|l|lcccccc|}
		\hline
		Model Class & \multicolumn{1}{l|}{Models}            & HR@5           & NDCG@5         & HR@10          & NDCG@10        & HR@20          & {NDCG@20}       \\ \hline
		& \multicolumn{1}{l|}{BPRMF \cite{rendle2009bpr}}                        & 38.72$\pm$0.10 & 29.66$\pm$0.12 & 47.53$\pm$0.10 & 32.50$\pm$0.07 & 57.21$\pm$0.22 & 34.95$\pm$0.13  \\
		& \multicolumn{1}{l|}{NeuMF \cite{he2017neural}}                        & 41.35$\pm$0.59 & 31.13$\pm$0.69 & 51.15$\pm$0.43 & 34.31$\pm$0.64 & 60.93$\pm$0.34 & 36.78$\pm$0.59  \\
		& \multicolumn{1}{l|}{SocialMF \cite{jamali2010matrix}}                     & 41.78$\pm$0.16 & 32.57$\pm$0.29 & 50.01$\pm$0.18 & 35.23$\pm$0.29 & 58.23$\pm$0.14 & 37.31$\pm$0.25  \\
		\multirow{-4}{*}{\begin{tabular}[c]{@{}l@{}}Matrix \\ Factorization \\ based\end{tabular}}
		& \multicolumn{1}{l|}{SoRec \cite{ma2008sorec}}                       & 40.81$\pm$0.33 & 31.14$\pm$0.30 & 49.61$\pm$0.16 & 33.99$\pm$0.24 & 58.42$\pm$0.19 & 36.22$\pm$0.25  \\ \hline
% 		& \multicolumn{1}{l|}{\textcolor{black}{MAGNN} \cite{fu2020magnn}}                           & -$\pm$- & -$\pm$- & -$\pm$- & -$\pm$- & -$\pm$- & -$\pm$-  \\
% 		& \multicolumn{1}{l|}{\textcolor{black}{DHGCN} \cite{manchanda2021schema}}                   & -$\pm$- & -$\pm$- & -$\pm$- & -$\pm$- & -$\pm$- & -$\pm$-  \\
		& \multicolumn{1}{l|}{\textcolor{black}{R-GCNs} \cite{schlichtkrull2018modeling}}            & 32.98$\pm$0.44 & 23.95$\pm$0.27 & 43.28$\pm$0.66 & 27.27$\pm$0.34 & 56.06$\pm$0.37 & 30.48$\pm$0.27  \\
		& \multicolumn{1}{l|}{\textcolor{black}{HAN} \cite{wang2019heterogeneous}}                   & 35.52$\pm$1.04 & 24.45$\pm$0.69 & 46.84$\pm$0.96 & 28.12$\pm$0.67 & 58.04$\pm$0.36 & 30.95$\pm$0.52  \\
		& \multicolumn{1}{l|}{\textcolor{black}{HGT} \cite{hu2020heterogeneous}}                     & 42.00$\pm$0.21 & 31.61$\pm$0.14 & 51.72$\pm$0.08 & 34.76$\pm$0.10 & 60.09$\pm$0.14 & 36.88$\pm$0.02  \\
		& \multicolumn{1}{l|}{GraphRec \cite{fan2019graph}}                     & 39.50$\pm$0.35 & 30.16$\pm$0.27 & 48.94$\pm$0.42 & 33.21$\pm$0.21 & 58.87$\pm$0.29 & 35.72$\pm$0.20  \\
		& \multicolumn{1}{l|}{LightGCN \cite{he2020lightgcn}}                     & 42.59$\pm$0.07 & 32.20$\pm$0.09 & 51.92$\pm$0.08 & 35.22$\pm$0.07 & 60.54$\pm$0.09 & 37.41$\pm$0.08  \\
		\multirow{-8}{*}{\begin{tabular}[c]{@{}l@{}}Graph Neural \\ Network based\end{tabular}}
		& \multicolumn{1}{l|}{DGRec \cite{song2019session}}                        & 40.36$\pm$0.25 & 30.52$\pm$0.16 & 49.67$\pm$0.14 & 33.53$\pm$0.15 & 59.26$\pm$0.19 & 35.95$\pm$0.15 \\ \hline
		& \multicolumn{1}{l|}{DMAN \cite{Tan_Zhang_Liu_Huang_Yang_Zhou_Hu_2021}}                     & 35.15$\pm$0.27 & 27.06$\pm$0.33 & 45.01$\pm$0.06 & 30.23$\pm$0.24 & 55.85$\pm$0.27 & 32.98$\pm$0.30 \\
		& \multicolumn{1}{l|}{TransRec \cite{he2017translation}}                     & 44.79$\pm$0.12 & 36.09$\pm$0.21 & 52.51$\pm$0.11 & 38.58$\pm$0.17 & 60.98$\pm$0.11 & 40.72$\pm$0.07 \\
		& \multicolumn{1}{l|}{SASRec \cite{kang2018self}}                          & 43.32$\pm$0.20 & 33.97$\pm$0.20 & 51.88$\pm$0.20 & 36.74$\pm$0.20 & 60.31$\pm$0.20 & 38.87$\pm$0.18  \\
		\multirow{-3}{*}{\begin{tabular}[c]{@{}l@{}}Sequence \\ based\end{tabular}}
		& \multicolumn{1}{l|}{ASAS \cite{manotumruksa2020sequential}}                         & 44.97$\pm$0.34 & 35.59$\pm$0.29 & 53.44$\pm$0.29 & 38.33$\pm$0.27 & 61.41$\pm$0.29 & 40.35$\pm$0.28  \\ 
% 		& \multicolumn{1}{l|}{STEN \cite{li2021extracting}}                         & 44.97$\pm$0.34 & 35.59$\pm$0.29 & 53.44$\pm$0.29 & 38.33$\pm$0.27 & 61.41$\pm$0.29 & 40.35$\pm$0.28  \\
		\hline
		Ours 
		& \multicolumn{1}{l|}{TEA-A}                      & 47.84$\pm$0.04 & 38.40$\pm$0.41 & 55.99$\pm$0.04 & 41.04$\pm$0.41 & 63.51$\pm$0.29 & 42.95$\pm$0.38  \\
		& \multicolumn{1}{l|}{TEA-S}                & \textbf{48.13}$\pm$0.25 & \textbf{38.65}$\pm$0.18 & \textbf{56.10}$\pm$0.17 & \textbf{41.24}$\pm$0.15 & \textbf{63.58}$\pm$0.08 & \textbf{43.13}$\pm$0.11  \\
		\hline 
	\end{tabular}}
\end{table*}

\subsection{Implementation Details}
We use PyTorch to implement our model and deploy it on RTX 2080 GPU. Hyper-parameter settings for all three datasets are as follows: embedding dimension $d=64$, batch size $B=1024$, dropout rate $p_\text{drop}=0.5$, L2 regularization weight $\gamma$=5e-4, negative sampling size $n_s=50$, sequence truncation length $L_s=50$, neighbor truncation length $L_n=20$, and learning rate $\eta=0.01$. We train all the methods with five different random seeds and report the means and standard deviations.  

\subsection{Evaluation Metrics}
We evaluate all the models with two widely used Top-N metrics: Hit Rate@$K$ (HR@$K$) and Normalized Discounted Cumulative Gain@$K$ (NDCG@$K$).
HR measures the percentage that recommended items contain at least one correct item interacted by the user, while NDCG considers the positions of correct recommended items. 
In the context of sequential recommendation, since we only test on the latest item in a user behavior sequence, HR is identical to recall and proportional to precision \cite{kang2018self}. 

Since it is time-consuming to rank all items for each user during the evaluation, we followed the strategy in \cite{kang2018self}. Specifically, for each user, we randomly sample 100 negative items and rank these items with the ground-truth item. HR and NDCG are estimated based on the ranking results. We report the experiment results for $K=5/10/20$.

\subsection{Compared Methods}
We compare our proposed models (TEA-S and TEA-A) based on TEA framework with three kinds of baselines: the matrix factorization based models, the graph neural networks based models, and the sequence recommendation methods. \\
\textbf{Matrix Factorization based Methods}:
\begin{itemize}
	\item BPRMF\cite{rendle2009bpr}: A general learning framework for personalized ranking recommendation uses implicit feedback. 
	\item NeuMF\cite{he2017neural}: It replaces the inner product with a multilayer perception (MLP) to learn the user-item interaction function.
	\item SocialMF \cite{jamali2010matrix}: It considers the social information and propagation of social information into the matrix factorization model.
	\item SoRec\cite{ma2008sorec}: It performs co-factorization on the user-item rating matrix and user-user social relations matrix.
\end{itemize}
\textbf{Graph Neural Network based Methods}:
\begin{itemize}
	\item GraphRec\cite{fan2019graph}: It uses the graph neural network to combine user behavior information and social network information into the recommendation task. For fairness, we discard the opinion/rate embedding in our implementation. 
	%\item MAGNN\cite{fu2020magnn}: The state-of-the-art heterogeneous graph embedding method. It uses the attention mechanism to aggregating different types of metapath and achieves user and item embeddings.  
	\item LightGCN \cite{he2020lightgcn}: A state-of-the-art graph-based collaborative filtering method. It explicitly integrates a bipartite graph structure into the embedding learning process to model the high-order connectivity in the user-item interaction graph.% In our experiments, we stack three layers of LightGCN. 
	\item DGRec \cite{song2019session}: A session-based recommendation method that combines the user action-temporal information and the social information via recurrent neural networks and dynamic graph attention networks.
% 	\item \textcolor{black}{MAGNN \cite{fu2020magnn}: The metapath aggregated graph neural networks leverage the intra-metapaths and inter-metapath to generate node embeddings and address the link prediction task.}
% 	\item \textcolor{black}{DHGCN \cite{manchanda2021schema}: The deep heterogeneous Graph Convolutional Network use the schema of heterogeneous to address the link prediction task.}
	\item \textcolor{black}{R-GCNs \cite{schlichtkrull2018modeling} use a different projection matrix for each relation in the heterogeneous graph and address the link prediction task.}
	\item \textcolor{black}{HAN \cite{wang2019heterogeneous} uses node-level and semantic-level hierarchical attention to model heterogeneous graph networks.}
	\item \textcolor{black}{HGT \cite{hu2020heterogeneous} designs node- and edge-type dependent parameters to characterize the heterogeneous attention.}
\end{itemize}
\textbf{Sequential Recommendation Methods:}
\begin{itemize}
	\item TransRec\cite{he2017translation}: A sequential recommendation method that models each user as a translation vector to capture the transition from the current item to the next. %In our experiments, we use squared L2 distance for the predictor. 
	\item SASRec \cite{kang2018self}: It leverages the Transformer\cite{vaswani2017attention} to capture users’ sequential behaviors. %We stack two self-attention blocks in our implementation. 
	\item ASASRec \cite{manotumruksa2020sequential}: An improved version of SASRec with an adversarial training strategy. 
	\item DMAN \cite{Tan_Zhang_Liu_Huang_Yang_Zhou_Hu_2021}: It effectively utilizes the sequential data by segmenting the overall behavior sequence and maintaining the long-term interests of users. 
% 	\item STEN \cite{li2021extracting}: We further consider the sequence recommendation that considers the social temporal informaiton. The STEN extracts the fine-grained user behavior sequence and the temporal networks.
\end{itemize}
\textbf{Model Variants:}
\begin{itemize}
	\item TEA-S: We use the GraphSAGE based aggregation method in the bipartite graph aggregation. 
	\item TEA-A: We use the Graph Attention mechanism based aggregation method in the bipartite graph aggregation.
	\item TEA-RS: We remove the time-restricted aggregation and use the GraphSAGE based aggregation method in the bipartite graph aggregation. 
	\item TEA-RA: We remove the time-restricted aggregation and use the Graph Attention mechanism based aggregation method in the bipartite graph aggregation. 
\end{itemize}

\subsection{Results}

\begin{table*}[htb]
	\caption{The performance evaluation of the compared methods on Yelp dataset. The values presented are averaged over 5 replicated with different random seeds.}
% 	\small
	\label{tab:yelp}
	\centering
	\scalebox{1.1}{
	\begin{tabular}{|l|lcccccc|}
		\hline
		Model Class & \multicolumn{1}{l|}{Models}            & HR@5           & NDCG@5         & HR@10          & NDCG@10        & HR@20          & {NDCG@20}       \\ \hline
		& \multicolumn{1}{l|}{BPRMF \cite{rendle2009bpr}}                        & 66.33$\pm$0.27 & 52.46$\pm$0.16 & 76.51$\pm$0.26 & 55.77$\pm$0.16 & 84.59$\pm$0.22 & 57.82$\pm$0.15  \\
		& \multicolumn{1}{l|}{NeuMF \cite{he2017neural}}                        & 70.38$\pm$0.26 & 56.14$\pm$0.28 & 79.35$\pm$0.12 & 59.06$\pm$0.24 & 86.14$\pm$0.12 & 60.79$\pm$0.22  \\
		& \multicolumn{1}{l|}{SocialMF \cite{jamali2010matrix}}                     & 64.82$\pm$0.24 & 49.69$\pm$0.24 & 76.27$\pm$0.28 & 53.42$\pm$0.21 & 84.99$\pm$0.28 & 55.63$\pm$0.19  \\
		\multirow{-4}{*}{\begin{tabular}[c]{@{}l@{}}Matrix \\ Factorization \\ based\end{tabular}}
		& \multicolumn{1}{l|}{SoRec \cite{ma2008sorec}}                        & 70.41$\pm$0.10 & 54.55$\pm$0.10 & 81.45$\pm$0.04 & 58.15$\pm$0.07 & 89.03$\pm$0.04 & 60.08$\pm$0.06  \\ \hline
% 		& \multicolumn{1}{l|}{\textcolor{black}{MAGNN} \cite{fu2020magnn}}                     & -$\pm$- & -$\pm$- & -$\pm$- & -$\pm$- & -$\pm$- & -$\pm$-  \\
% 		& \multicolumn{1}{l|}{\textcolor{black}{DHGCN} \cite{manchanda2021schema}}             & -$\pm$- & -$\pm$- & -$\pm$- & -$\pm$- & -$\pm$- & -$\pm$-  \\
		& \multicolumn{1}{l|}{\textcolor{black}{R-GCNs} \cite{schlichtkrull2018modeling}}      & 79.90$\pm$0.09 & 63.77$\pm$0.14 & 89.17$\pm$0.06 & 66.80$\pm$0.13 & 94.45$\pm$0.05 & 68.15$\pm$0.12  \\
		& \multicolumn{1}{l|}{\textcolor{black}{HAN} \cite{wang2019heterogeneous}}             & 74.00$\pm$0.62 & 56.46$\pm$0.30 & 85.35$\pm$0.51 & 60.16$\pm$0.26 & 92.35$\pm$0.37 & 61.95$\pm$0.23  \\
		& \multicolumn{1}{l|}{\textcolor{black}{HGT} \cite{hu2020heterogeneous}}               & 76.60$\pm$0.11 & 61.19$\pm$0.01 & 85.79$\pm$0.12 & 64.18$\pm$0.01 & 92.15$\pm$0.12 & 65.80$\pm$0.01  \\
		& \multicolumn{1}{l|}{GraphRec \cite{fan2019graph}}                     & 68.37$\pm$0.23 & 51.44$\pm$0.27 & 81.55$\pm$0.17 & 55.74$\pm$0.18 & 90.61$\pm$0.17 & 58.05$\pm$0.16  \\
		& \multicolumn{1}{l|}{LightGCN \cite{he2020lightgcn}}                     & 73.04$\pm$0.21 & 57.10$\pm$0.21 & 84.39$\pm$0.07 & 60.80$\pm$0.19 & 92.08$\pm$0.07 & 62.76$\pm$0.17  \\
		\multirow{-8}{*}{\begin{tabular}[c]{@{}l@{}}Graph Neural \\ Network based\end{tabular}}
		& \multicolumn{1}{l|}{DGRec \cite{song2019session}}                        & 76.22$\pm$0.24 & 60.18$\pm$0.28 & 86.57$\pm$0.18 & 63.55$\pm$0.26 & 92.93$\pm$0.08 & 65.18$\pm$0.16 \\ \hline
		& \multicolumn{1}{l|}{DMAN \cite{Tan_Zhang_Liu_Huang_Yang_Zhou_Hu_2021}}                     & 72.93$\pm$0.33 & 57.45$\pm$0.16 & 83.64$\pm$0.34 & 60.94$\pm$0.29 & 91.03$\pm$0.25 & 62.82$\pm$0.26 \\
		& \multicolumn{1}{l|}{TransRec \cite{he2017translation}}                     & 75.81$\pm$0.15 & 60.63$\pm$0.16 & 80.19$\pm$0.20 & 64.00$\pm$0.15 & 93.13$\pm$0.12 & 65.78$\pm$0.15 \\
		& \multicolumn{1}{l|}{SASRec \cite{kang2018self}}                          & 69.28$\pm$0.39 & 53.18$\pm$0.43 & 81.66$\pm$0.08 & 57.21$\pm$0.37 & 90.36$\pm$0.08 & 59.43$\pm$0.34  \\
		\multirow{-3}{*}{\begin{tabular}[c]{@{}l@{}}Sequence \\ based\end{tabular}}
		& \multicolumn{1}{l|}{ASASRec \cite{manotumruksa2020sequential}}                         & 72.97$\pm$0.13 & 56.76$\pm$0.10 & 84.53$\pm$0.04 & 60.53$\pm$0.09 & 92.18$\pm$0.04 & 62.48$\pm$0.07  \\ 
% 		& \multicolumn{1}{l|}{STEN \cite{li2021extracting}}                         & 44.97$\pm$0.34 & 35.59$\pm$0.29 & 53.44$\pm$0.29 & 38.33$\pm$0.27 & 61.41$\pm$0.29 & 40.35$\pm$0.28  \\
\hline
		Ours
% 		& \multicolumn{1}{l|}{TEA-A} & 80.38$\pm$0.25 & 65.42$\pm$0.36 & \textbf{88.99}$\pm$0.14 & 68.23$\pm$0.33 & \textbf{94.11}$\pm$0.10 & 69.54$\pm$0.21 \\
% 		& \multicolumn{1}{l|}{TEA-S} & \textbf{80.43}$\pm$0.18 & \textbf{65.59}$\pm$0.26 & 88.97$\pm$0.08 & \textbf{68.37}$\pm$0.23 & 94.09$\pm$0.07 & \textbf{69.68}$\pm$0.21 \\ 
		& \multicolumn{1}{l|}{TEA-A} & 81.13$\pm$0.25 & 66.79$\pm$0.36 & 88.65$\pm$0.14 & 69.24$\pm$0.33 & 93.50$\pm$0.10 & 70.43$\pm$0.21 \\
		& \multicolumn{1}{l|}{TEA-S} & \textbf{84.08}$\pm$0.18 & \textbf{70.16}$\pm$0.26 & \textbf{90.57}$\pm$0.08 & \textbf{72.29}$\pm$0.23 & \textbf{94.68}$\pm$0.07 & \textbf{73.33}$\pm$0.21 \\ 
		
		\hline 
	\end{tabular}}
\end{table*}

\begin{table*}[htb]
	\caption{The performance evaluation of the compared methods on WeChat dataset. The values presented are averaged over 5 replicated with different random seeds.}
% 	\small
	\label{tab:Wechat}
	\centering
	\scalebox{1.1}{
	\begin{tabular}{|l|ccccccc|}
		\hline
		Model Class & \multicolumn{1}{l|}{Models}            & HR@5            & NDCG@5          & HR@10           & NDCG@10          & HR@20          & {NDCG@20}        \\ \hline
		& \multicolumn{1}{l|}{BPRMF \cite{rendle2009bpr}}                        & 62.33$\pm$0.12 & 56.38$\pm$0.12 & 68.55$\pm$0.18 & 58.38$\pm$0.14 & 75.60$\pm$0.10 & 60.16$\pm$0.14 \\
		& \multicolumn{1}{l|}{NeuMF \cite{he2017neural}}                        & 68.58$\pm$0.16 & 61.66$\pm$0.16 & 75.26$\pm$0.18 & 63.82$\pm$0.16 & 82.53$\pm$0.14 & 65.65$\pm$0.15 \\
		& \multicolumn{1}{l|}{SocialMF \cite{jamali2010matrix}}                     & 68.25$\pm$0.13 & 61.30$\pm$0.42 & 74.62$\pm$0.07 & 63.36$\pm$0.39 & 81.44$\pm$0.04 & 65.09$\pm$0.37 \\
		\multirow{-4}{*}{\begin{tabular}[c]{@{}l@{}}Matrix \\ Factorization \\ based\end{tabular}}
		& \multicolumn{1}{l|}{SoRec \cite{ma2008sorec}}                       & 73.66$\pm$0.04 & 66.43$\pm$0.11 & 79.60$\pm$0.02 & 68.36$\pm$0.10 & 85.56$\pm$0.03 & 69.86$\pm$0.09 \\ \hline
% 			& \multicolumn{1}{l|}{\textcolor{black}{MAGNN} \cite{fu2020magnn}}                    & - & - & - & - & - & - \\
% 		& \multicolumn{1}{l|}{\textcolor{black}{DHGCN} \cite{manchanda2021schema}}                & - & - & - & - & - & - \\
		& \multicolumn{1}{l|}{\textcolor{black}{R-GCNs} \cite{schlichtkrull2018modeling}}         & 77.57$\pm$0.18 & 66.43$\pm$0.25 & 85.04$\pm$0.15 & 68.85$\pm$0.24 & 91.04$\pm$0.07 & 70.38$\pm$0.22 \\
		& \multicolumn{1}{l|}{\textcolor{black}{HAN} \cite{wang2019heterogeneous}}                & - & - & - & - & - & - \\
		& \multicolumn{1}{l|}{\textcolor{black}{HGT} \cite{hu2020heterogeneous}}                  & 74.53$\pm$0.11 & 65.70$\pm$0.10 & 81.45$\pm$0.06 & 67.94$\pm$0.08 & 88.02$\pm$0.03 & 69.61$\pm$0.08 \\
		& \multicolumn{1}{l|}{GraphRec \cite{fan2019graph}}                     & 66.04$\pm$0.33 & 52.17$\pm$0.29 & 76.80$\pm$0.25 & 55.66$\pm$0.26 & 85.72$\pm$0.18 & 57.93$\pm$0.24 \\
		& \multicolumn{1}{l|}{LightGCN \cite{he2020lightgcn}}                     & - & - & - & - & - & - \\
		\multirow{-8}{*}{\begin{tabular}[c]{@{}l@{}}Graph Neural \\ Network based\end{tabular}}
		& \multicolumn{1}{l|}{DGRec \cite{song2019session}}                        & 74.99$\pm$0.22 & 63.94$\pm$0.24 & 82.29$\pm$0.14 & 66.31$\pm$0.24 & 88.52$\pm$0.09 & 67.89$\pm$0.20 \\ \hline
		& \multicolumn{1}{l|}{DMAN \cite{Tan_Zhang_Liu_Huang_Yang_Zhou_Hu_2021}}                     & - & - & - & - & - & - \\
		& \multicolumn{1}{l|}{TransRec \cite{he2017translation}}                     & 73.54$\pm$0.17 & 64.87$\pm$0.16 & 80.06$\pm$0.14 & 66.99$\pm$0.15 & 86.03$\pm$0.14 & 68.50$\pm$0.14 \\
		& \multicolumn{1}{l|}{SASRec \cite{kang2018self}}                          & 76.37$\pm$0.35 & 65.97$\pm$0.35 & 83.76$\pm$0.21 & 68.37$\pm$0.29 & 89.94$\pm$0.11 & 69.94$\pm$0.23 \\
		\multirow{-3}{*}{\begin{tabular}[c]{@{}l@{}}Sequence \\ based\end{tabular}}
		& \multicolumn{1}{l|}{ASASRec \cite{manotumruksa2020sequential}}                         & 78.28$\pm$0.23 & 68.13$\pm$0.25 & 85.19$\pm$0.21 & 70.38$\pm$0.19 & 90.81$\pm$0.14 & 71.80$\pm$0.13 \\ 
        \hline
		Ours
		& \multicolumn{1}{l|}{TEA-A}                      & 82.06$\pm$0.25 & 73.47$\pm$0.26 & 87.61$\pm$0.16 & 75.28$\pm$0.21 & 92.18$\pm$0.17 & 76.44$\pm$0.13 \\
		& \multicolumn{1}{l|}{TEA-S}                & \textbf{83.23}$\pm$0.19 & \textbf{76.12}$\pm$0.21 & \textbf{88.12}$\pm$0.13 & \textbf{77.70}$\pm$0.18 & \textbf{92.42}$\pm$0.17 & \textbf{78.79}$\pm$0.16 \\
		\hline 
	\end{tabular}}
% 	\vspace{-3mm}
\end{table*}

% Mean hr5=0.7499, hr10=0.8229, hr20=0.8852, ndcg5=0.6394, ndcg10=0.6631, ndcg20=0.6789	
% Std  hr5=0.0022, hr10=0.0014, hr20=0.0009, ndcg5=0.0024, ndcg10=0.0021, ndcg20=0.0020

% Table \ref{tab:epin}, \ref{tab:yelp} and \ref{tab:Wechat} present the recommendation performance of all methods on the three datasets, respectively. 

% \subsubsection{Ablation Analysis}

{Tables \ref{tab:epin}, \ref{tab:yelp} and \ref{tab:Wechat} present the recommendation performance of all the methods on the three datasets, respectively. We do not report the performance of LightGCN, DMAN, and HAN on WeChat Official Accounts dataset because of the limitation of GPU memory.}

First, by modeling social influence, the performances of social-aware methods (SocialMF, SoRec, GraphRec, and DGRec) are improved compared with that of BPRMF in most cases, which is consistent with previous works. This {observation} indicates that social information reflects users' interests effectively. 
Second, the sequence based methods (DGRec, TransRec, SAS, and ASAS) also perform comparably well. These improvements reflect the importance of temporal information on recommendation tasks.  
Third, DGRec and our proposed methods (including TEA-S and TEA-A) that combine social information and temporal information achieve much better performance, especially on large datasets. 
At last, our proposed TEA-S and TEA-A consistently outperform all the {compared methods} on both public and real-world datasets with an average improvement of 3.15\% on HR@10 and 8.38\% on NDCG@10 against the best {competitor}. The significant improvements validate the effectiveness of aggregating the user behavior sequence and the influence between the users. We also observe that performance of TEA-A is slightly {lower} than that of TEA-S, indicating that the graph attention mechanism is difficult to handle the high sparsity of temporally evolving heterogeneous graphs. %The simple mean pooling aggregation of TEA-GraphSAGE provides robustness to our model. 

\subsection{Ablation Study}
\begin{figure*}[htbp]
\centering
\begin{minipage}[t]{0.32\textwidth}
\centering
\label{fig:ablation_epinion}
\includegraphics[width=\textwidth]{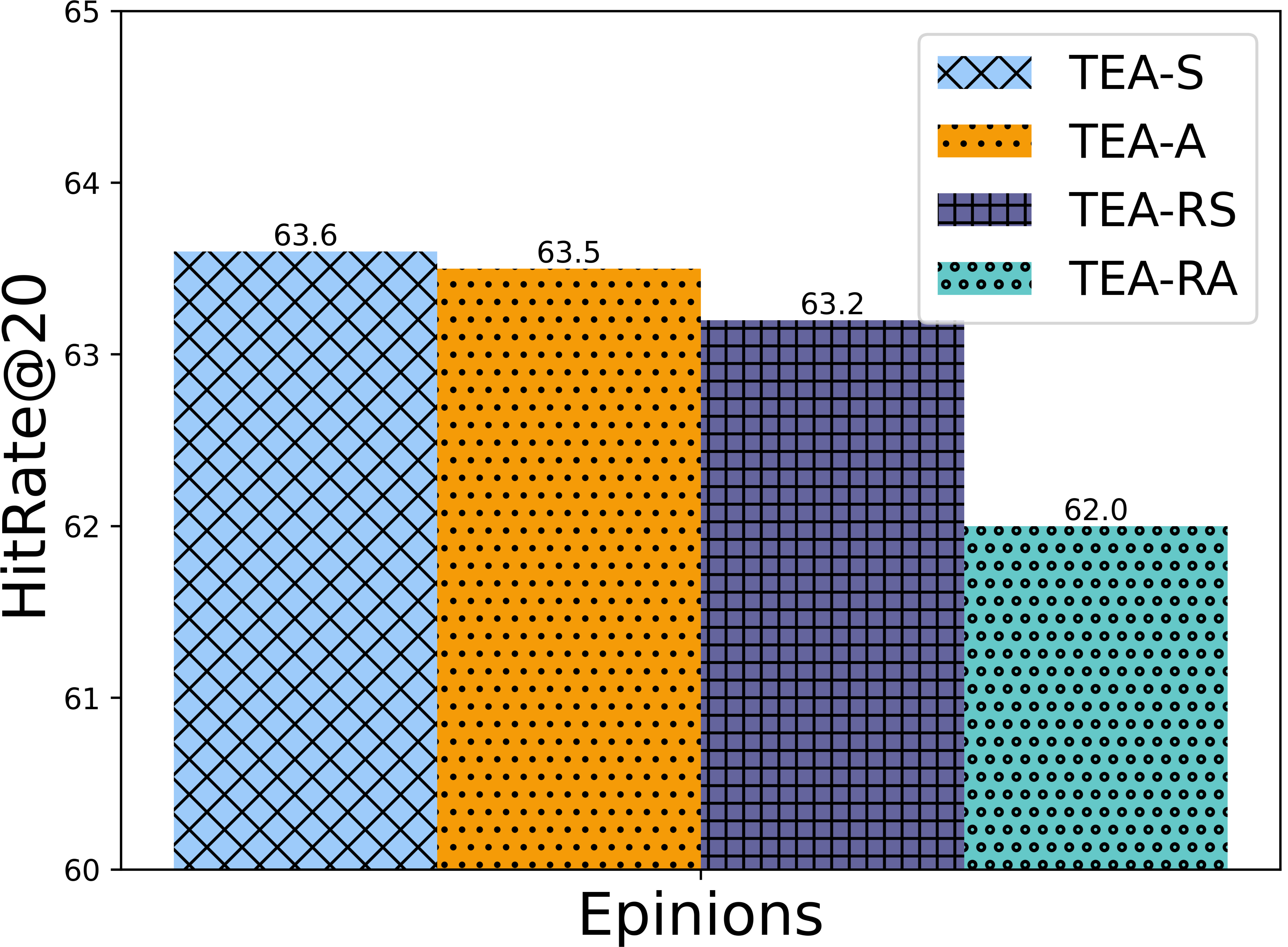}
\caption{The Experiment results of ablation studies on Epinions.}
\end{minipage}
\begin{minipage}[t]{0.32\textwidth}
\centering
\label{fig:ablation_yelp}
\includegraphics[width=\textwidth]{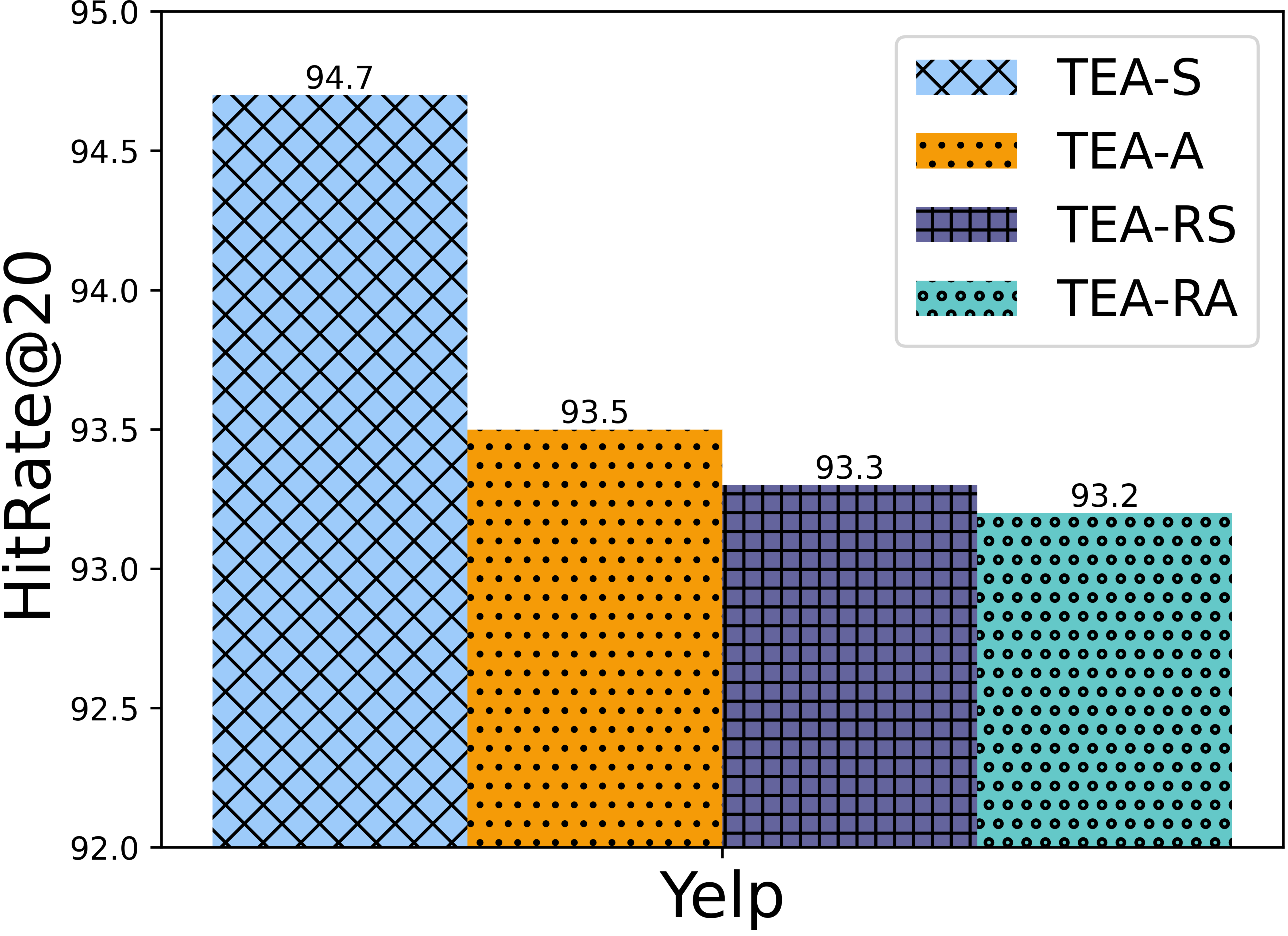}
\caption{The Experiment results of ablation studies on Yelp.}
\end{minipage}
\begin{minipage}[t]{0.32\textwidth}
\centering
\label{fig:ablation_epinion}
\includegraphics[width=\textwidth]{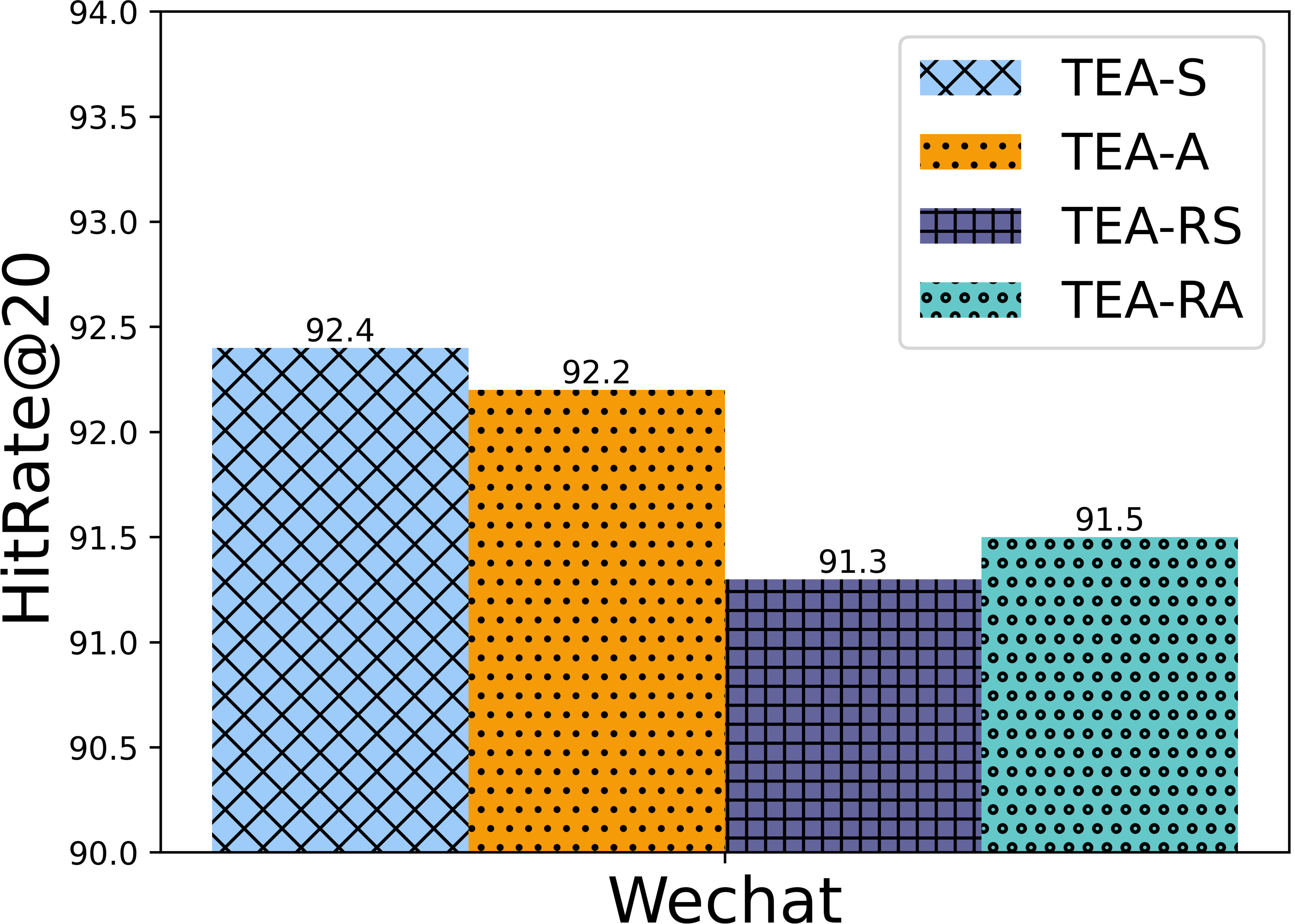}
\caption{The Experiment results of ablation studies on Wechat.}
\end{minipage}
\end{figure*}
In order to evaluate the effectiveness of the time-restricted aggregation, we remove the aforementioned aggregation module and obtain the variants \textbf{TEA-RS} and \textbf{TEA-RA}. {experimental results on each dataset are} shown in {Figures} 3, 4, and 5, respectively. {From these results}, we can find that the models with time-restricted aggregation achieve a better performance, especially the results on the Yelp dataset. We also find that the promotion in Epinions dataset is not so remarkable, this is {since} the social networks in Epinions are much denser than that of Yelp. To some extent, the experiment results reflect that the proposed time-restricted aggregation can mitigate the drawbacks of sparse social networks and user-item interactions.

\subsection{The Sensitivity of Hyper-parameters}

\begin{figure}[t]
\centering
\label{fig:hypm_emb}
\begin{minipage}[htbp]{0.24\textwidth}
\centering
\includegraphics[width=\textwidth]{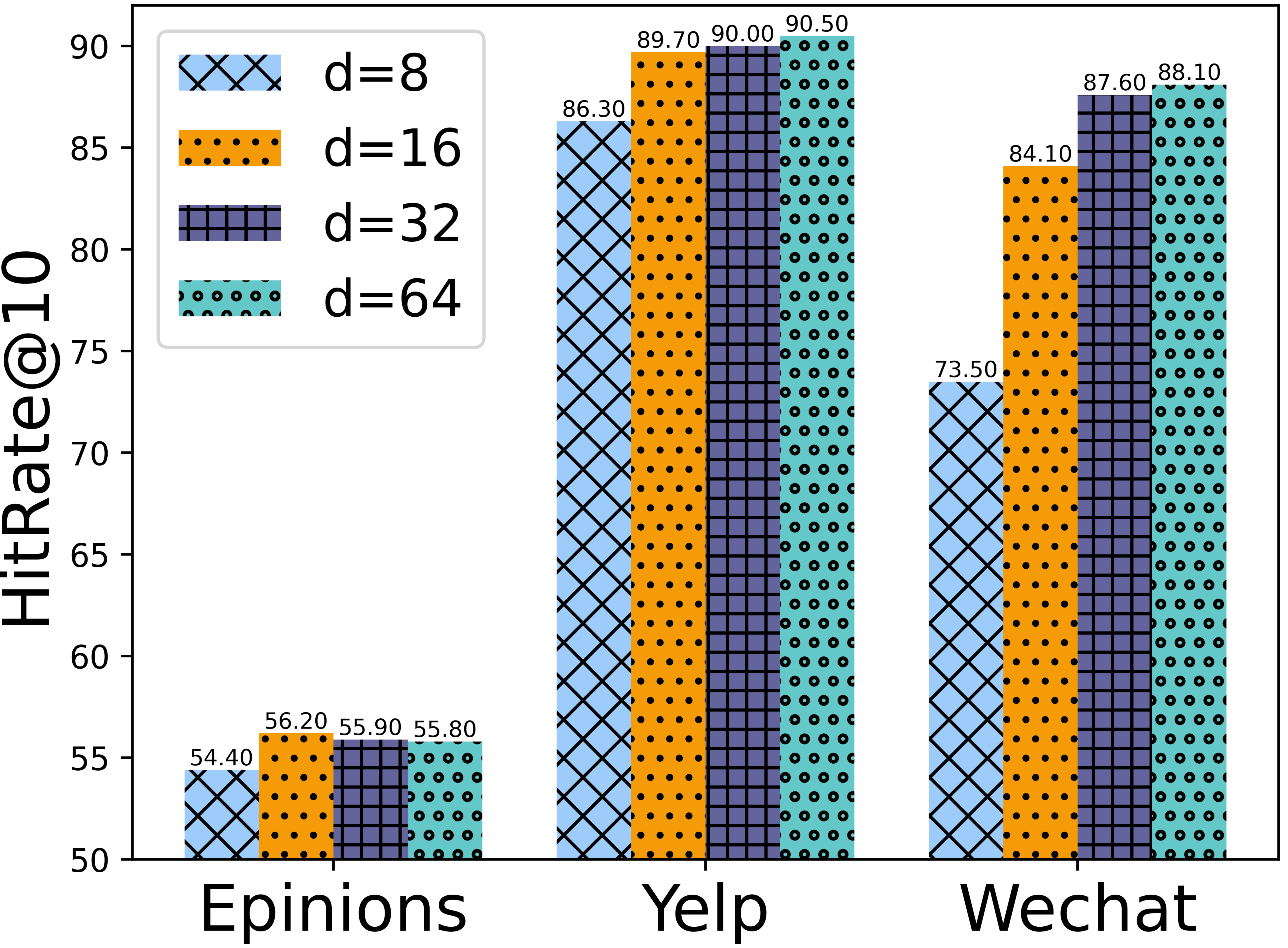}
\end{minipage}
\begin{minipage}[htbp]{0.24\textwidth}
\centering
\includegraphics[width=\textwidth]{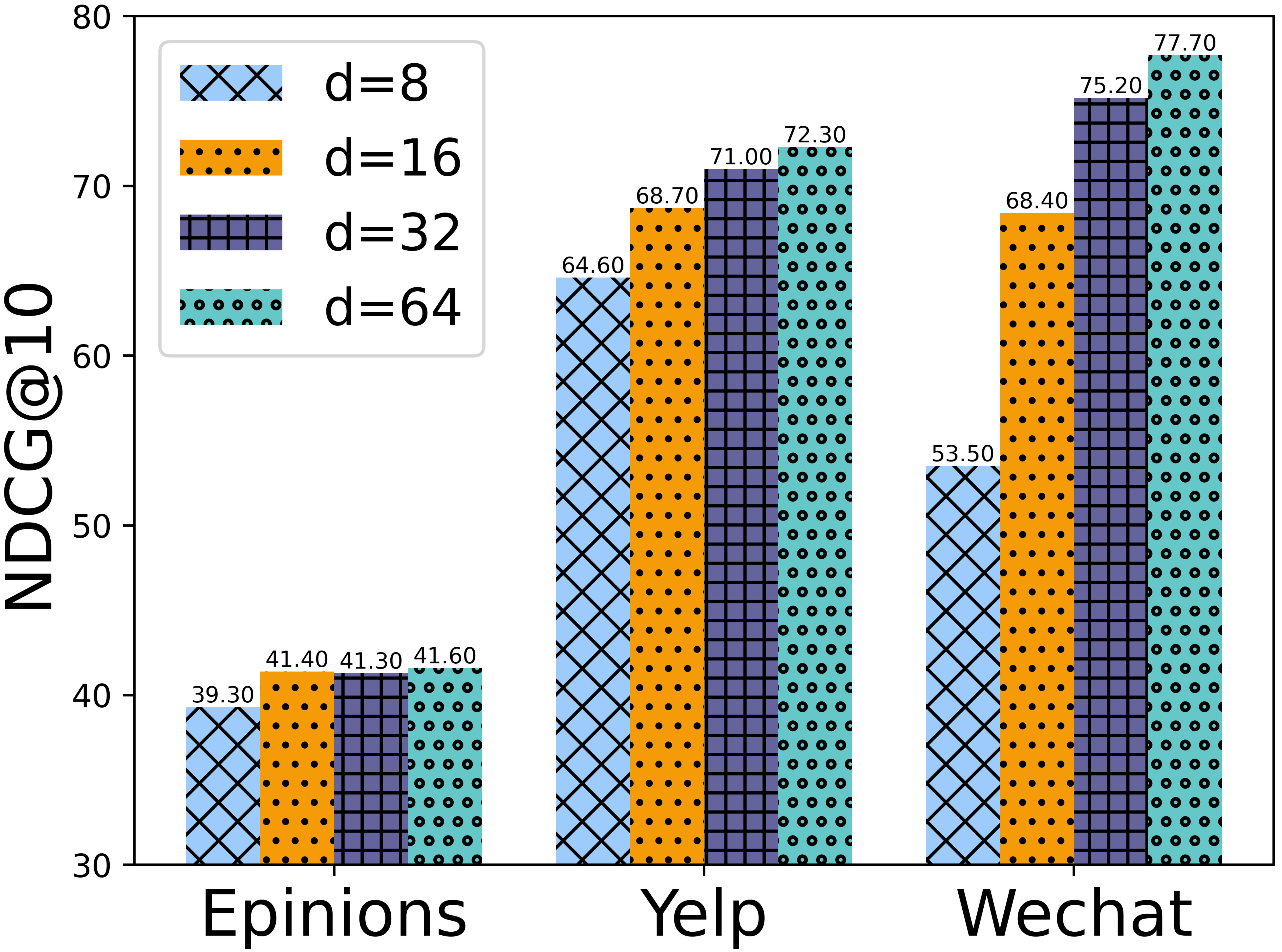}
\end{minipage}
\caption{The sensitivity of the embedding dimension $d$.}
\end{figure}

\begin{figure}[t]
\centering
\begin{minipage}[htbp]{0.240\textwidth}
\centering
\includegraphics[width=\textwidth]{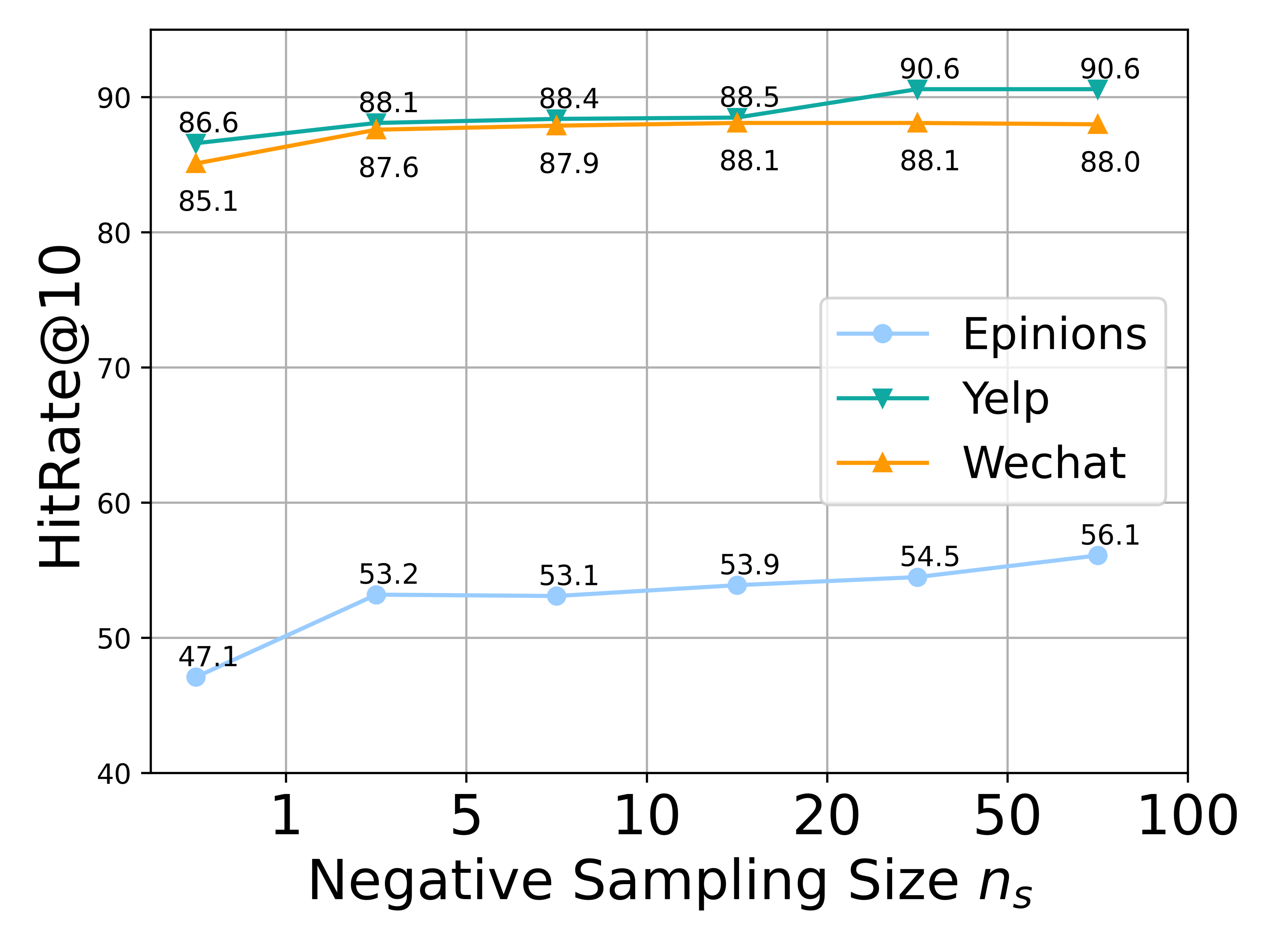}
\caption{The sensitivity of the negative sampling size $n_s$.}
\end{minipage}
\begin{minipage}[htbp]{0.240\textwidth}
\centering
\includegraphics[width=\textwidth]{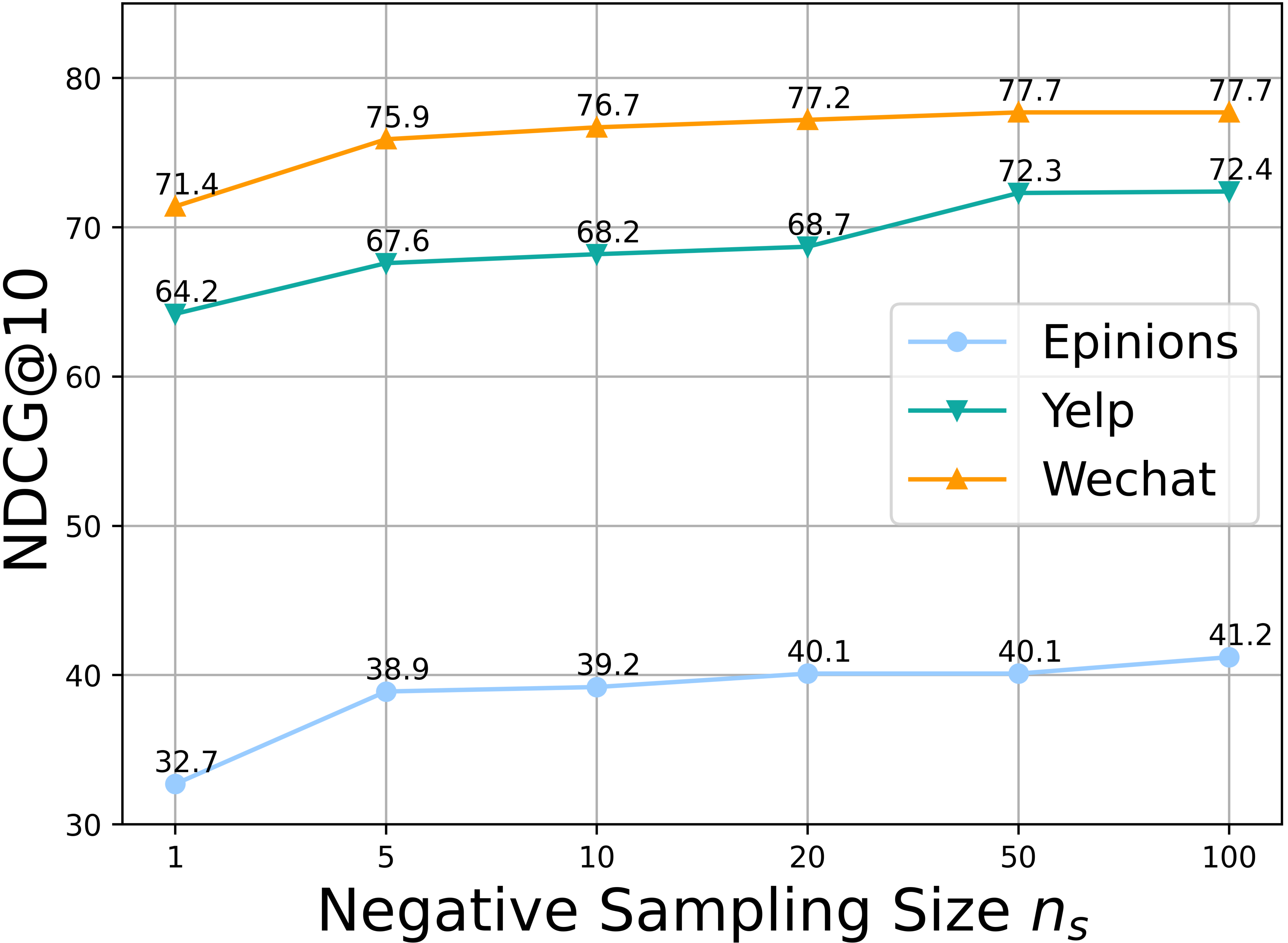}
\caption{The sensitivity of the negative sampling size $n_s$.}
\end{minipage}
\end{figure}

\subsubsection{Embedding Dimension.}

According to the experiment results shown in {Figure} 6, we analyze the sensitivity of the embedding dimension $d$ by showing HR@10 and NDCG@10 of our proposed {TEA-S} with $d$ varying from 8 to 64 \textcolor{black}{on the Epinions, Yelp and Wechat dataset}. \textcolor{black}{According to the experiment results, we find that the experiment results ($d=32$ and $64$) of HR@10 on Epinions are slightly lower than the result of $d=16$, but the experiment results of NDCG@10 on the other datasets reflect that larger dimension benefits the model performance and a small dimension ($d=16$) is enough for TEA-S to achieve the best performance.}
% \blue{For Epinion with a smaller size, the embedding dimension $d=16$ is enough for {TEA-S} to achieve the best performance.}

% \subsubsection{Number of Negative Samples.}
% \begin{figure}[t]
% 	\centering
% 	\label{fig:hypm_ns}
% 	\scalebox{0.48}
% 	{\includegraphics[width=\textwidth]{fig/hypm_ns}}
% 	\caption{The sensitivity of the negative sampling size $n_s$.}
% \end{figure}

% \subsubsection{Number of Negative Samples.}
% \begin{figure}[htbp]
% 	\centering
% 	\caption{The sensitivity of the negative sampling size $n_s$.}
% 	\label{fig:hypm_ns}
% 	\scalebox{0.45}
% 	{\includegraphics[width=\textwidth]{fig/hypm_ns}}
% \end{figure}
\subsubsection{Sensitivity of the Number of Negative Samples.}
{Figures} 7 and 8 show the sensitivity of the number of negative samples $n_s$ in Equation (\ref{equ:final_loss}) by showing HR@10 and NDCG@10 of our proposed {TEA-S} with $n_s$ varying from 1 to 100 on \textcolor{black}{Epinions, Yelp and Wechat dataset}. The variant with $n_s=5$ performs comparably well, though using $n_s \geq 10$ still boosts performance especially on the large-scale dataset, which means that using more negative samples is helpful to estimate the item transition probability. 
The variant with $n_s=100$ achieves a similar performance to the default setting $n_s=50$, which indicates that our model is stable with $n_s$.

% \begin{figure}[t]
% 	\centering
% 	\scalebox{0.48}
% 	{\includegraphics[width=\textwidth]{fig/transition.pdf}}
% 	\caption{The item category transition matrix on train and test dataset of Epinions. }
% 	\label{fig:heat}
% \end{figure}
% \vspace{-2ex}

% \subsection{Visualization}
% As shown in Figure \ref{fig:heat}, we visualize the category based transition matrix of Epinions dataset \green{since the number of items is too large}. Please note that the categories are not used at training. Figure \ref{fig:heat} (a) is the transition matrix we calculate on the training set, and (b) is the transition matrix estimated from the output of our model on the test set. We can find that these two transition matrices are similar, which reveals that our proposed method learns the relationship among items to some extend.
% \blue{more explanations}

\section{Conclusion}\label{sec:conclu}
This paper presents a temporally evolving aggregations framework for the sequential recommendation. Beginning from the original conditional random field, we derive the unified objective function for the sequential recommendation, \textcolor{black}{which leverages the social influence between users and the dynamic user-item heterogeneous graph.} The proposed framework provides the insights and principles for designing the sequential recommendation model. We further provide two different implementations of the proposed framework. Experimental results on three real-world datasets show that the \textbf{TEA} framework outperforms state-of-the-art methods.

\section{Acknowledgments}

% This research was supported in part by Natural Science Foundation of China (61876043, 61976052), Science and Technology Planning Project of Guangzhou (201902010058). 
We would like to thank Lingling Yi and Li Li from WeChat for their help and supports on this work.

% if have a single appendix:
%\appendix[Proof of the Zonklar Equations]
% or
%\appendix  % for no appendix heading
% do not use \section anymore after \appendix, only \section*
% is possibly needed

% use appendices with more than one appendix
% then use \section to start each appendix
% you must declare a \section before using any
% \subsection or using \label (\appendices by itself
% starts a section numbered zero.)
%

% you can choose not to have a title for an appendix
% if you want by leaving the argument blank

% use section* for acknowledgment
% \ifCLASSOPTIONcompsoc
%   % The Computer Society usually uses the plural form
%   \section*{Acknowledgments}
% \else
%   % regular IEEE prefers the singular form
%   \section*{Acknowledgment}
% \fi

% The authors would like to thank...

% Can use something like this to put references on a page
% by themselves when using endfloat and the captionsoff option.
\ifCLASSOPTIONcaptionsoff
  \newpage
\fi

% trigger a \newpage just before the given reference
% number - used to balance the columns on the last page
% adjust value as needed - may need to be readjusted if
% the document is modified later
%\IEEEtriggeratref{8}
% The "triggered" command can be changed if desired:
%\IEEEtriggercmd{\enlargethispage{-5in}}

% references section

% can use a bibliography generated by BibTeX as a .bbl file
% BibTeX documentation can be easily obtained at:
% http://mirror.ctan.org/biblio/bibtex/contrib/doc/
% The IEEEtran BibTeX style support page is at:
% http://www.michaelshell.org/tex/ieeetran/bibtex/

\bibliographystyle{IEEEtran}
\bibliography{main_2}

% argument is your BibTeX string definitions and bibliography database(s)
%\bibliography{IEEEabrv,../bib/paper}
%
% <OR> manually copy in the resultant .bbl file
% set second argument of \begin to the number of references
% (used to reserve space for the reference number labels box)

% \begin{thebibliography}{1}

% \bibitem{IEEEhowto:kopka}
% H.~Kopka and P.~W. Daly, \emph{A Guide to \LaTeX}, 3rd~ed.\hskip 1em plus
%   0.5em minus 0.4em\relax Harlow, England: Addison-Wesley, 1999.

% \end{thebibliography}

% biography section
% 
% If you have an EPS/PDF photo (graphicx package needed) extra braces are
% needed around the contents of the optional argument to biography to prevent
% the LaTeX parser from getting confused when it sees the complicated
% \includegraphics command within an optional argument. (You could create
% your own custom macro containing the \includegraphics command to make things
% simpler here.)
%\begin{IEEEbiography}[{\includegraphics[width=1in,height=1.25in,clip,keepaspectratio]{mshell}}]{Michael Shell}
% or if you just want to reserve a space for a photo:
% \newpage
\vspace{-8ex}

% if you will not have a photo at all:
% \begin{IEEEbiographynophoto}{John Doe}
% Biography text here.
% \end{IEEEbiographynophoto}

% insert where needed to balance the two columns on the last page with
% biographies
%\newpage

% \begin{IEEEbiographynophoto}{Jane Doe}
% Biography text here.
% \end{IEEEbiographynophoto}

% You can push biographies down or up by placing
% a \vfill before or after them. The appropriate
% use of \vfill depends on what kind of text is
% on the last page and whether or not the columns
% are being equalized.

%\vfill

% Can be used to pull up biographies so that the bottom of the last one
% is flush with the other column.
%\enlargethispage{-5in}

% that's all folks
\end{document}